\documentclass[%
 reprint,
]{revtex4-2}
\usepackage[colorlinks,linkcolor=blue,urlcolor=blue,citecolor=blue]{hyperref}
\usepackage{graphicx}% Include figure files
\usepackage{dcolumn}% Align table columns on decimal point
\usepackage{bm}% bold math
\usepackage{subfigure}
\begin{document}

\preprint{APS/123-QED}

\title{High-repetition-rate seeded free-electron laser enhanced by self-modulation}% Force line breaks with \\

\author{Hanxiang Yang\textsuperscript{1,}\textsuperscript{2}}
\author{Jiawei Yan\textsuperscript{3}}
\author{Haixiao Deng\textsuperscript{4,}}%
\thanks{denghaixiao@zjlab.org.cn}
\affiliation{\textsuperscript{1}Shanghai Institute of Applied Physics, Chinese Academy of Sciences, Shanghai 201800, China\\
	\textsuperscript{2}University of Chinese Academy of Sciences, Beijing 100049, China\\
	\textsuperscript{3}European XFEL, 22869 Schenefeld, Germany\\
	\textsuperscript{4}Shanghai Advanced Research Institute, Chinese Academy of Sciences, Shanghai 201210, China
}%

\date{\today}% It is always \today, today,
\begin{abstract}
Recently, the self-modulation scheme of a weakly pre-bunched electron beam has been proposed [Yan \emph{et al.}, \href{https://doi.org/10.1103/PhysRevLett.126.084801}{Physical Review Letters \textbf{126}, 084801 (2021)}], which is of great promise for high-repetition-rate seeded free-electron lasers (FELs), such as high-gain harmonic generation (HGHG). In this paper, the self-modulation scheme is systematically analyzed and optimized, and further experiments in which the self-modulator is resonant at the second harmonic of the seed laser are conducted. The three-dimensional numerical simulations show that the required seed laser intensity in the self-modulation scheme is around three orders of magnitude lower than that of the standard HGHG through the optimization of the beam size or the peak current. More importantly, by reasonably setting the initial energy modulation and the resonance of the self-modulator, a more prominent bunching factor and lase at the 30th harmonic of the seed laser can be achieved in a single-stage HGHG. Moreover, the experiment results confirm that varying the resonance conditions of the self-modulator can still amplify the laser-induced energy modulation, where coherent radiation generated even at the 12th harmonic can be observed. These results indicate that the self-modulation scheme can remarkably reduce the requirements of the seed laser system while improving the harmonic up-conversion efficiency, which paves the way for realizing high-repetition-rate and fully coherent soft x-ray FELs.
\end{abstract}

\maketitle

%\tableofcontents
\section{\label{sec:sec1}Introduction}
High-gain free-electron lasers (FELs) can generate wavelength-tunable pulses of high brightness, opening a range of new research in various fields, including biology, chemistry, physics, and materials science \cite{Pellegrini2016,Huang2021}. However, most successful FEL-based experiments to date are based on the properties of single-shot pulses, such as internal structure studies of materials \cite{SERAFINI2019}. In contrast, fully coherent FELs with high repetition rates are critical for some photon-hungry experiments, such as the fine time-resolved analysis of matter with spectroscopy and photon scattering.

Currently, most of the x-ray FEL facilities worldwide \cite{Emma2010,Ishikawa2012,Kang2017,Decking2020,Prat2020} are based on the mechanism of self-amplified spontaneous emission (SASE) \cite{Kondratenko1980,Bonifacio1994}. The SASE scheme can obtain high-brightness pulses with a sub\r{a}ngstrom wavelength and a femtosecond duration. Because of the mode selection process, SASE pulses can become nearly transversely coherent. However, the longitudinal coherence of the SASE pulses is limited by the slippage of radiation within the FEL gain length. Self-seeding schemes \cite{FELDHAUS1997341,Gianluca2011} are proposed to improve the longitudinal coherence but at the cost of shot-to-shot intensity fluctuations.

Seeded FELs \cite{Yu1991,Yu1997,Stupakov2009,Xiang2009,Deng2013} that use a coherent external seed laser to trigger the initial amplification process of XFEL are ideal for obtaining fully coherent FEL pulses with a small pulse energy fluctuation. In a typical configuration of seeded FELs, such as the high-gain harmonic generation (HGHG) \cite{Yu2000}, sinusoidal energy modulation is introduced by an external seed laser in a modulator through the laser-beam interaction. Then, a magnetic chicane referred to as the dispersive section is used to convert the energy modulation into periodic density bunching comprising frequency components at high harmonics. Finally, the microbunched beams are sent through a relatively long undulator referred to as the radiator, in which the coherent radiation of the harmonic of interest is generated and further amplified through the FEL process. In addition, sophisticated phase-space manipulation techniques, such as the echo-enabled harmonic generation (EEHG) \cite{Stupakov2009,Xiang2009} and phase-merging enhanced harmonic generation (PEHG) \cite{Deng2013}, have been developed to improve the frequency multiplication efficiency of the single stage. Both analytical calculations and experimental results demonstrate that seeded FELs can produce high-power soft x-ray radiation with a narrow bandwidth using an ultraviolet seed laser \cite{Allaria2012,Allaria2013,Liu2013,Zhao2012,Hemsing2016,RebernikRibic2019,Feng2019}.

In recent years, the high-repetition-rate operation of XFEL has been proposed to obtain pulses with high average brightness, which have enormous potential in a variety of scientific studies \cite{SERAFINI2019,Decking2020,Stohr2011,Zhao:FEL2017}. Based on superconducting accelerators, FLASH \cite{Ackermann2007} and the European XFEL \cite{Decking2020} can reach repetition rates of 1 MHz and 4.5 MHz in a burst mode, respectively. Moreover, SHINE \cite{Zhao:FEL2017} and LCLS-II \cite{Stohr2011} are designed to achieve 1 MHz in a continuous wave mode. A seeded FEL with a repetition rate of 1 MHz can meet the requirements of specific experiments, such as high-resolution spectroscopy techniques. However, it is currently impossible for state-of-the-art laser systems to obtain laser pulses with sufficient peak power for the energy modulation of the electron beam and at the same time have such a high repetition rate.

Different methods have been proposed to realize high-repetition-rate seeded FELs. The utilization of an FEL oscillator in the hard x-ray or far ultraviolet frequency range as the seeding source for subsequent cascades has been proposed \cite{Gandhi2013,Li2017}. In addition, a resonator-like seed recirculation feedback system is proposed to reuse the radiation in the modulator section to modulate the following electron bunches \cite{Ackermann2020,Petrillo202000,Paraskaki2021}. The use of a relatively longer modulator has been suggested to reduce the peak power requirements of the external seed laser \cite{duning2011,Paraskaki202100,Wang2022}. Recently, the self-modulation of the electron beam has been experimentally demonstrated to be an effective method of reducing the requirement for an external seed laser \cite{Yan2021,yan20212}. Moreover, further studies have shown that the self-modulation method has similar stability to HGHG and has the potential to achieve very high harmonics \cite{Paraskaki202100,zhao2022}. More analysis of various methods to high-repetition-rate seeded FELs has been analyzed \cite{Jia2021}.

In this paper, we first systematically analyze the crucial parameters in the self-modulation scheme to provide design and optimization directions for its further application in seeded FELs, such as HGHG. The schematic layout and principle of the self-modulation HGHG are presented in Sec.~\ref{sec:sec2}. In order to demonstrate the feasibility of the self-modulation HGHG for high-repetition-rate and fully coherent soft x-ray FELs, intensive numerical simulations based on the parameters of the Shanghai soft x-ray FEL facility (SXFEL) are carried out in Sec.~\ref{sec:sec3} and Sec.~\ref{sec:sec4}, respectively. We further verify that the self-modulation scheme can generate higher harmonic bunching after changing the resonance conditions of the self-modulator, and the experimental results are shown in Sec.~\ref{sec:sec5}. The summary and prospects are given in Sec.~\ref{sec:sec6}.

\section{\label{sec:sec2}The self-modulation scheme}
The standard HGHG typically adopts a UV seed laser pulse to imprint a longitudinal sinusoidal energy modulation on the electron beam in a short modulator, with a periodicity of the seed laser wavelength. The following dispersive chicane is used for the associated longitudinal density modulation, which can be defined as the bunching factor. The $n$-th harmonic bunching factor can be written as \cite{Yu1991}
\begin{equation}
b_{n}=\left|J_{n}(nAB)\right| \exp \left(-\frac{n^{2} B^{2}}{2}\right),
\label{eq:1}
\end{equation}
where $A =\Delta \gamma/\sigma_\gamma$, $\gamma$ is the beam energy in units of $mc^2$, $\Delta \gamma$ is the energy modulation amplitude, and $\sigma_\gamma$ is the uncorrelated energy spread. $B = k R_{56}\sigma_\gamma/\gamma$ is the dimensionless dispersion parameter, $R_{56}$ is the dispersion strength of the chicane, $k$ is the wavenumber of the seed laser, $n$ is the harmonic number, and $J_n$ is the $n$-th order first class Bessel function. Equation (\ref{eq:1}) shows that the bunching factor decreases exponentially with the rise of harmonic number $n$. To obtain the maximum bunching factor at the $n$-th harmonic of the seed laser, a large $A$ is required, and $B$ should be optimized simultaneously. Generally, the $\Delta \gamma$ should be $n$ times larger than $\sigma_\gamma$. However, the energy spread effects induced by the seed laser limits its access to the higher harmonic number, e.g., larger than the FEL pierce parameter $\rho$ when the FEL process cannot gain exponentially in the following radiators. The total energy spread $\sigma_{\gamma^{'}}$ of the electron beam can be written as $\sigma_{\gamma^{'}}=\sqrt{\sigma_{\gamma}^{2}+\Delta \gamma^{2}/2}$ \cite{Yu2002}. From the notation of Ref. \cite{Huang2007}, the FEL pierce parameter is derived as
\begin{equation}
  \rho=\left[\frac{1}{8 \pi} \frac{I}{I_{A}}\left(\frac{K[J J]_{1}}{1+K^{2} / 2}\right)^{2}
  \frac{\gamma \lambda^{2}}{2 \pi \sigma_{x}^{2}}\right]^{1 / 3},
  \label{eq:2}
\end{equation}
where $K$ is the undulator parameter, $[JJ]_1$ is the usual Bessel function factor associated with a planar undulator, $\lambda$ is the FEL resonant wavelength, $I_{A}$ is the Alfvén current, and $I$ is the peak current of the electron beam, and $\sigma_{x}$ is the rms transverse beam size. Consequently, the maximum harmonic up-conversion of the standard HGHG is typically limited to 15 under the constraint of satisfying both large energy modulation and the requirement on high-gain FELs. In order to obtain shorter wavelengths, a cascading multi-stage scheme can be adopted, in which the FEL output of the first stage HGHG is used as the seed laser for the next \cite{Yu1997}. The cascaded HGHG is more sensitive to beam imperfections, which is challenging to operate at the shorter wavelength in practice \cite{Allaria2013,Liu2013}.
\begin{figure}[b]
	\includegraphics[width=0.4\textwidth]{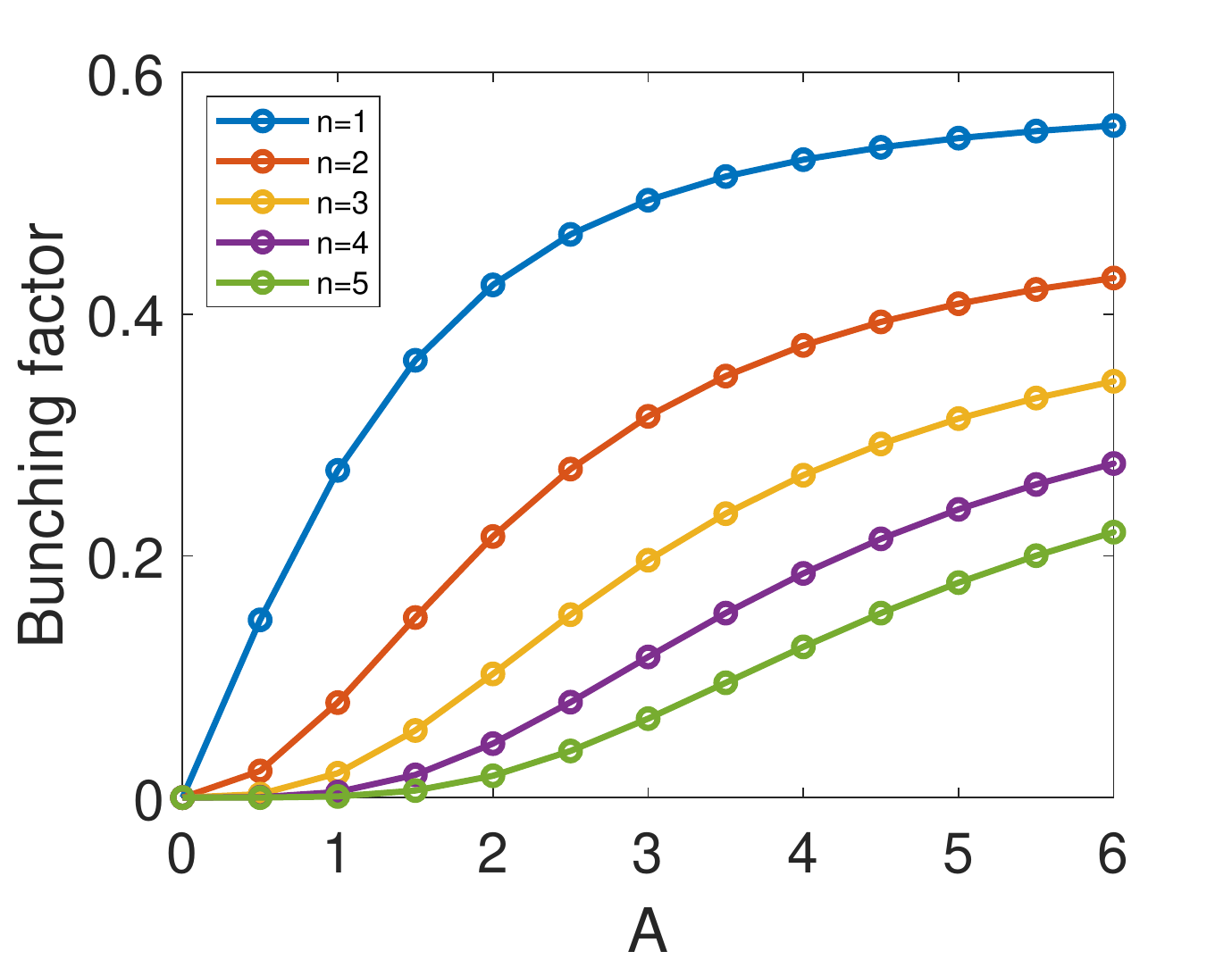}% Here is how to import EPS art
	\caption{\label{fig:AvsBf}Bunching factor versus energy modulation amplitude $A$ at various harmonic numbers. Each line corresponds to the maximum bunching factor at the optimal dispersion strength.}
\end{figure}

Furthermore, the bunching factor versus $A$ at various harmonic numbers is illustrated in Fig.~\ref{fig:AvsBf}, according to Eq. (\ref{eq:1}). Under a particular $A$, each line represents the maximum achievable bunching factor at the optimal $R_{56}$. For instance, inducing a relatively small energy modulation $A$ = 2, the maximum bunching factor of the fundamental, second, and third harmonics are 0.43, 0.22, and 0.10, respectively. The fundamental wavelength contains a significant bunching factor under this weak energy modulation, which means that the further amplification can produce coherent energy modulation and reduce the requirement for the seed laser power towards high-repetition-rate FELs \cite{Yan2021}. Besides, it can be shown that if $A$ increases, such as to 4, the bunching factors of the fundamental, second, and third harmonics increase to 0.55, 0.37, and 0.27, respectively. The bunching factor of the fundamental and second harmonics increases by 28\% and 68\%, respectively, while the third harmonic increases by a factor of 1.7. Therefore, the induced relatively weak initial energy modulation can be further amplified in the second, third, or higher harmonics towards short-wavelength seeded FELs \cite{zhao2022}.
\begin{figure}[tb]
	\includegraphics[width=0.48\textwidth]{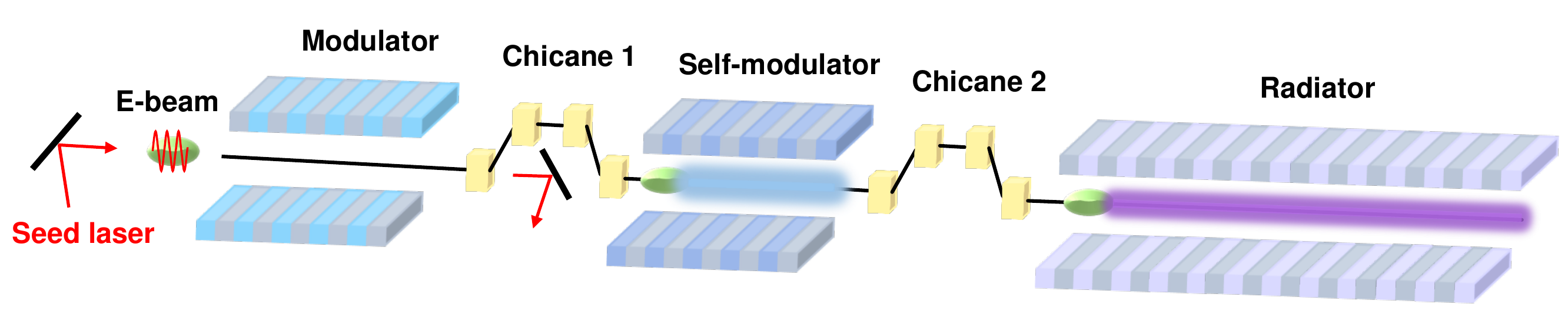}% Here is how to import EPS art
	\caption{\label{Fig:SelfHGHG}Schematic layout of the self-modulation HGHG setup. A self-modulation HGHG includes extra dispersive chicane and self-modulator, further amplifying laser-induced energy modulation to obtain a higher harmonic bunching factor.}
\end{figure}

The schematic layout of the self-modulation HGHG setup is shown in Fig.~{\ref{Fig:SelfHGHG}}. Compared with the standard HGHG, the self-modulation HGHG includes an extra chicane and a short undulator, the so-called self-modulator, which is used to reduce the initial requirements on the seed laser power significantly. A UV seed laser is utilized to interact with the electron beam in the first modulator and then generate a relatively weak energy modulation. After passing through the first chicane, the energy modulation of the electron beam is converted into longitudinal density modulation so-called pre-bunched beam. Then the laser-induced energy modulation is further amplified in the self-modulator. Moreover, optimizing the dispersion strength of the second chicane can achieve a more prominent bunching factor at a higher harmonic of the seed laser in the following amplifier. It should be noted that the external seed laser is separated from the dispersive section between the modulators. This scheme is well compatible with existing seeding lines of FEL facilities, because it is similar to the EEHG setup. The energy modulation amplitude induced in the first modulator $A_1 =\Delta \gamma_1/\sigma_\gamma$ and enhanced in the self-modulator $A_2 =\Delta \gamma_2/\sigma_\gamma$ should be analyzed and optimized with two dispersive chicanes’ $R_{56}^1$ and $R_{56}^2$ simultaneously. The enhanced coherent energy modulation amplitude is calculated with the energy spread in the exit of self-modulator. According to the theory of seeded FELs \cite{Yu1991}, the gain process consists of three regimes: the CHG with quadratic growth, exponential growth, and saturation. The CHG radiation can be generated in the self-modulator whose total length is around two gain lengths. $L_{g} \equiv \lambda_{u}/{4 \pi \sqrt{3} \rho}$, where $\lambda_{u}$ is the undulator period, and $\rho$ is the FEL pierce parameter in Eq. (\ref{eq:2}). Here, $L_{g}$ is the one-dimensional (1D) gain length with vanishing energy spread. In the CHG regime, the harmonic electric field grows linearly along the self-modulator length $z$, and the power grows quadratically along $z$. At the entrance of the self-modulator, the pre-bunched beam contains high harmonic components. Thereby, the self-modulator can enhance the CHG radiation. Assuming the electron beam is a longitudinal uniform distributed, the CHG power can be written as \cite{Yu2002}
\begin{equation}
  P_{\mathrm{coh}}=\frac{Z_{0}\left(K[J J]_{1} \eta L_{g} I b_{m1}\right)^{2}}{32 \pi \sigma_{x}^{2} \gamma^{2}},
\label{eq:3}
\end{equation}
where $Z_{0}$ = 377 $\Omega$ is the vacuum impedance, $\eta = L_{m2}/L_{g}$ is the modulator scaling factor, $L_{m2}$ is the length of the self-modulator, $b_{m1}$ is the bunching factor on the resonant wavelength of the self-modulator obtained from the pre-bunched beam, and the other variables follow the same notation as in Eq. (\ref{eq:2}).
Generally, assuming a Gaussian external seed laser interacts with the electron beam in the modulator undulator, we can obtain the amplitude of the energy modulation as (see, e.g., \cite{Huang2004})
\begin{equation}
  \Delta \gamma(r)=\sqrt{\frac{P}{P_{0}}} \frac{K[JJ]_{1} 
  L_{u}}{\gamma \sigma_{r}} \exp({-\frac{r^{2}}{4 \sigma_{r}^{2}}}),
    \label{eq:4}
\end{equation}
where $P$ is the peak power of the external seed laser, $P_0 = I_{A}mc^2/e \approx 8.7$ GW, $r$ is the radial position of the electron beam, and $\sigma_{r}$ is the rms laser spot size in the modulator undulator. The laser spot size $\sigma_{r}$ should be as comparable as possible to the electron beam size $\sigma_{x}$ to obtain the maximum energy modulation $\Delta \gamma(0)$. However, to satisfy transverse overlap between the electron beam and external seed laser, the laser spot size should be larger than the beam size, typically a few hundred micrometers. The energy modulation process of the self-modulator is similar to that of the conventional modulator with an external seed laser, as it is the coherent energy modulation process generated by the CHG radiation that requires a modification to Eq. (\ref{eq:4}). The electric field strength of CHG radiation grows linearly with the self-modulator length. Assuming that the electric field strength at the exit of the self-modulator is $E_{\mathrm{coh}}$, the coherent energy modulation process of the self-modulator can be regarded as modulated by the seed laser with an electric field strength of $E_{\mathrm{coh}}/2$ corresponding to the peak power of $P_{\mathrm{coh}}/4$. Meanwhile, for simplicity, we can assume that the spot size of the CHG radiation is larger than the beam size in the self-modulator and that both are independent of each other. The energy modulation in the self-modulator can be simplified. Thus, the modified maximum energy modulation $\Delta \gamma$ can be rewritten as
\begin{equation}
  \Delta \gamma=\sqrt{\frac{P_{\mathrm{coh}}}{4P_{0}}} \frac{K[JJ]_{1}L_{u}}{\gamma \sigma^{'}_{r}}
  \label{eq:5}
\end{equation}
where $\sigma^{'}_{r}$ is the equivalent spot size of the CHG radiation.

Here, the critical parameters of the self-modulation HGHG is described for design and optimization. With Eq. (\ref{eq:4}) and Eq. (\ref{eq:5}), it is clear that $A_{2}\propto\sqrt{P_{\mathrm{coh}}}$, $A_{1}\propto\sqrt{P_{\mathrm{seed}}}$, where $P_{\mathrm{seed}}$ is the external seed laser power in the first modulator. As shown in Fig.~\ref{fig:AvsBf}, since the $A$ is generally larger than 2, the harmonic bunching factor can be roughly considered proportional to the square root of $A$. The harmonic bunching factor $b_{m2}$ at the entrance of radiators can be simplified as $b_{m2}\propto\sqrt{A_{2}}$. The $A_1$ is typically lower than 2, which can be roughly estimated as proportional to the bunching factor $b_{m1}\propto A_{1}$. Consequently, the scaling harmonic bunching factor can be easily derived as
\begin{equation}
  b_{m2}\propto\frac{\eta^{1/2}I^{1/3}P^{1/4}_{\mathrm{seed}}}{\sigma_{x}^{1/6}}.
   \label{eq:6}
\end{equation}

Equation (\ref{eq:6}) illustrates the scaling relationship between the harmonic bunching factor and critical parameters such as modulator scaling factor $\eta$, peak current $I$, peak power of the external seed laser $P_{\mathrm{seed}}$, and the electron beam size $\sigma_x$, which can prior estimate the seed laser power of the self-modulation HGHG under different electron beam and undulator characteristics. For example, in the nominal case with beam size $\sigma_{x0}$, peak current $I_0$, seed laser power $P_{\mathrm{seed0}}$, and modulator scaling factor $\eta$ = 2, then the $b_{m2}$ is obtained. When the beam size is reduced by half of $\sigma_{x0}/2$ and the peak current remains $I_0$, producing the same bunching factor as $b_{m2}$, the seed laser power is only $P_{\mathrm{seed}}/4$, corresponding to $\eta$ = 3.18; for comparison, when the peak current is doubled to 2$I_0$ and the beam size remains $\sigma_{x0}$, producing the same $b_{m2}$, the seed laser power is also reduced to $P_{\mathrm{seed}}/4$, corresponding to $\eta$ = 2.52. Note that the scaling bunching factor contains three preconditions: (1) the modulator scaling factor should generally be less than or approximately equal to 3, according to Eq. (\ref{eq:3}); (2) the spot size of CHG radiation in the self-modulator is independent of the beam size, according to Eq. (\ref{eq:5}); (3) the energy modulation amplitude of the first modulator is weak, typically less than 2. Therefore, during design and optimization of the self-modulation HGHG, the seed laser power requirement can be significantly reduced by optimizing the peak current and beam size, and this process is substantially achieved by reducing the gain length in the self-modulator. In practice, the length of the modulator undulator cannot be adjusted arbitrarily, thereby optimizing the electron beam quality and the FODO cell of the modulators is a more feasible method for the self-modulation scheme.
% \begin{equation}
%   \eta=\frac{L_{m2}}{L_{g}}
%   \label{eq:13}
% \end{equation}
\begin{table}
	\caption{\label{tab:table1}Main electron beam parameters of the SXFEL-UF.}
	\begin{ruledtabular}
		\begin{tabular}{lcc}
			Parameters				&Value	&Unit\\ \hline
			{\textbf{\textit{Electron beam}}}&		&\\
			Energy					&1.4		&GeV\\
			Slice energy spread 	&50     	&keV\\
			Normalized emittance  	&1	        &mm·mrad\\
			Bunch charge			&600	    &pC\\
			Bunch length (FWHM)		&800    	&fs\\
			Peak current (Gaussian) &700	    &A\\
			Beam size (RMS)         &100        &$\mu$m\\
		\end{tabular}
	\end{ruledtabular}
\end{table}

\section{\label{sec:sec3}Towards high repetition rates}
To investigate the principle and limits of the self-modulation HGHG, numerical simulation were carried out by GENESIS \cite{REICHE1999} utilizing the main electron beam parameters of the SXFEL user facility (SXFEL-UF) listed in Table~\ref{tab:table1}. The SXFEL-UF is aimed to generate soft x-ray pulses towards the wavelength of the spectrum of 2.3 nm - 4.4 nm (“water window”), which contains the SASE line and seeding line \cite{Liu2021}. The single-stage EEHG and cascaded EEHG-HGHG are initially adopted in the baseline design. A 266-nm seed laser with peak power in the order of 100 MW is adopted. Towards high-repetition-rate FELs, the peak power requirements of external seed lasers should be further reduced to circumvent the limitations of the state-of-the-art laser system. In addition, it is currently challenging to achieve a harmonic up-conversion number of externally seeding schemes to above 100 towards short-wavelength FELs. We employ SXFEL-UF electron beam parameters as a nominal case to further demonstrate the performance and application potential of the self-modulation scheme in terms of high-repetition-rate and short-wavelength FELs.
\begin{table}
	\caption{\label{tab:table2}Main simulation parameters of the seed laser and the undulators.}
	\begin{ruledtabular}
		\begin{tabular}{lcc}
			Parameters				&Value	&Unit\\ \hline
			{\textbf{\textit{Seed laser}}}&		&\\
			Wavelength                  &266		&nm\\
			Peak power (Standard HGHG)  &17-75		&MW\\
			Peak power (Self-modulation HGHG)  &0.019 - 1.6		&MW\\
			Pulse duration (FWHM)    	&150     	&fs\\
			Rayleigh length           	&5	        &m\\
			Spot size (RMS)		    	&325	    &$\mu$m\\
			&&\\
			{\textbf{\textit{Modulator}}}&		&\\
			$K$                         &9.891		&\\
			Period                      &8		    &cm\\
			Length                      &1.6		&m\\
			&&\\
			{\textbf{\textit{Self-modulator}}}&		&\\
			$K$                         &5.593 - 9.891		&\\
			Period                      &8		    &cm\\
			Length                      &1.6 or 2		    &m\\
			&&\\
			{\textbf{\textit{Radiator}}}&		&\\
			$K$                         &1.823 - 4.239		&\\
			Period                     	&5     	&cm\\
			Length           	        &3	        &m\\
		\end{tabular}
	\end{ruledtabular}
\end{table}

\begin{table*}
	\caption{\label{tab:table3}The main working points of the standard HGHG and self-modulation HGHG from the 8th to 15th harmonics of the seed laser. The nominal case is listed in Table~\ref{tab:table1}, the second case is beam size of 50 $\mu$m and peak current of 700 A, and the third case is that of 100 $\mu$m and 1400 A, respectively.}
	\begin{ruledtabular}
		\begin{tabular}{lcccc}
			%			&\multicolumn{1}{c}{\textbf{Stage1 EEHG}}&\multicolumn{1}{c}{\textbf{Stage1 EEHG}}\\
			&Specification&The nominal case&The second case&The third case\\ \hline
			{\textbf{Standard HGHG}}
			&Seed laser power range (MW)&19 - 76             &17 - 72          &19 - 76\\
			&$R_{56}$ (mm)              &0.10 - 0.21         &0.11 - 0.20      &0.10 - 0.21\\
			{\textbf{Self-modulation HGHG}}
			&Seed laser power range (MW)    &0.080 - 1.6        &0.019 - 0.16      &0.022 - 0.28\\
			&$R_{56}^{1}$ (mm)              &0.67 - 1.17        &0.92 - 1.45       &0.62 - 1.24\\
			&$R_{56}^{2}$ (mm)              &0.048 - 0.15      &0.069 - 0.18      &0.052 - 0.16\\
		\end{tabular}
	\end{ruledtabular}
\end{table*}

As described in Sec.~\ref{sec:sec2}, the modulator scaling factor $\eta$, peak current $I$, seed laser power $P_{\mathrm{seed}}$, and the electron beam size $\sigma_x$ are critical for high-repetition-rate FELs. The beam size is reduced to 50 $\mu$m from the nominal case of 100 $\mu$m to compare the effect of beam size on the self-modulation HGHG. In addition, the peak current is increased to 1400 A for the nominal case of 700 A to compare the effect of the peak current further. It should be noted that the absolute length of the self-modulator of 1.6 m remains as the usable length of the self-modulator at SXFEL-UF, while modulator scaling factor $\eta$ as an intermediate variable can change accordingly.

To reasonably compare the effects of beam size and peak current on self-modulation HGHG efficiency, the same target bunching factor of 8\% is used in all cases of this section. The harmonic bunching factor is obtained at the entrance of radiators for the 8th - 15th harmonics of the seed laser. The parameters of the FODO lattice in the following radiators are kept at the same value. The steady-state simulations are carried out with the electron and undulator parameters listed in Table~\ref{tab:table1} and \ref{tab:table2} to optimize working points of the self-modulation HGHG fast, i.e., the seed laser power and dispersion strength. The numerical simulations of standard HGHG are given in contrast with the self-modulation HGHG, as summarized in Table~\ref{tab:table3}.
\begin{figure}
	\includegraphics[width=0.4\textwidth]{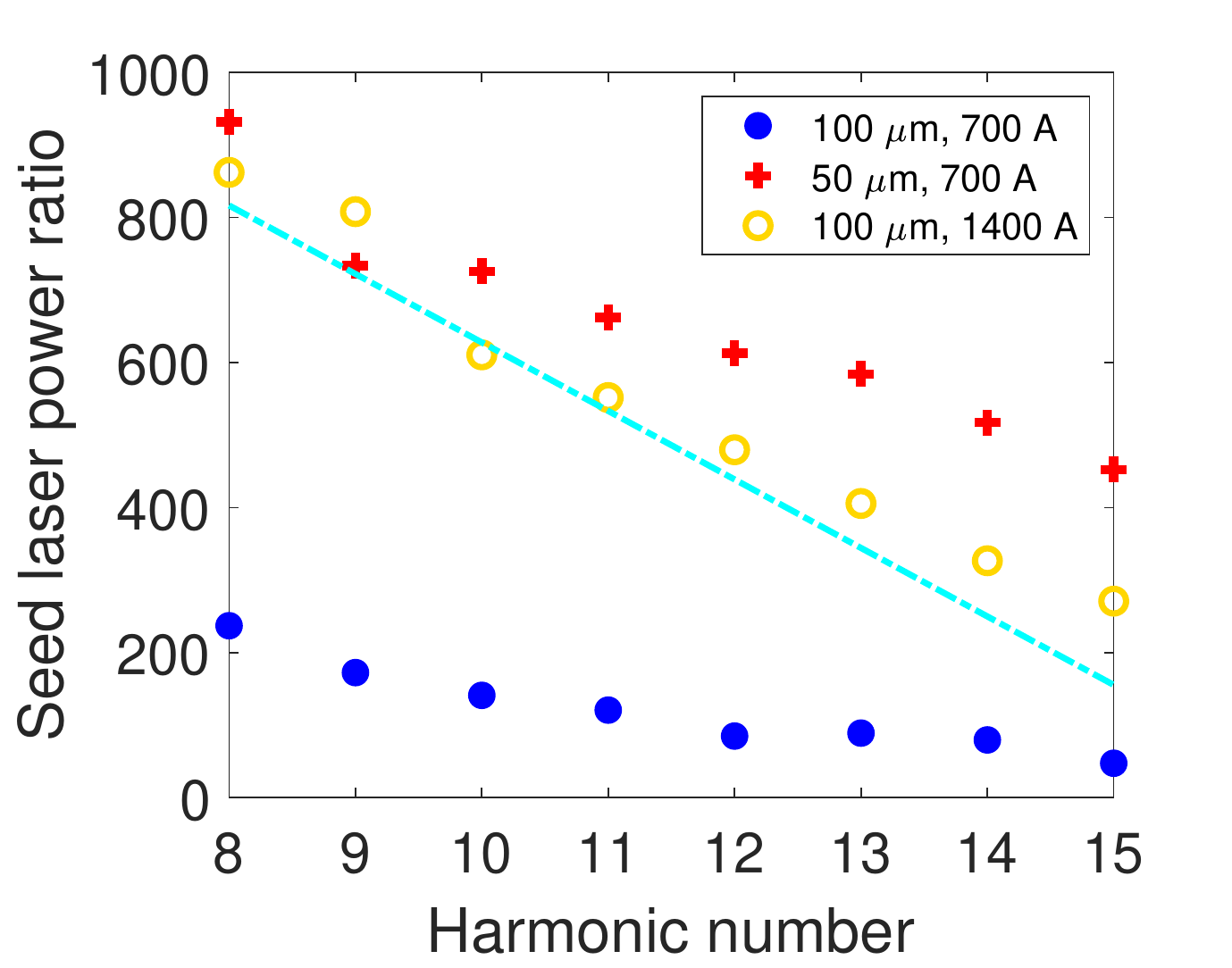}% Here is how to import EPS art
	\caption{\label{fig:3}The seed laser power ratio of the standard HGHG and self-modulation HGHG in different cases. The blue dot, red cross, and yellow circle correspond to the nominal case, the second case of beam size of 50 $\mu$m and peak current of 700 A, and the third case of 100 $\mu$m and 1400 A, respectively. The dotted line corresponds to the scaling curve.}
\end{figure}

\begin{figure}
	\includegraphics[width=0.4\textwidth]{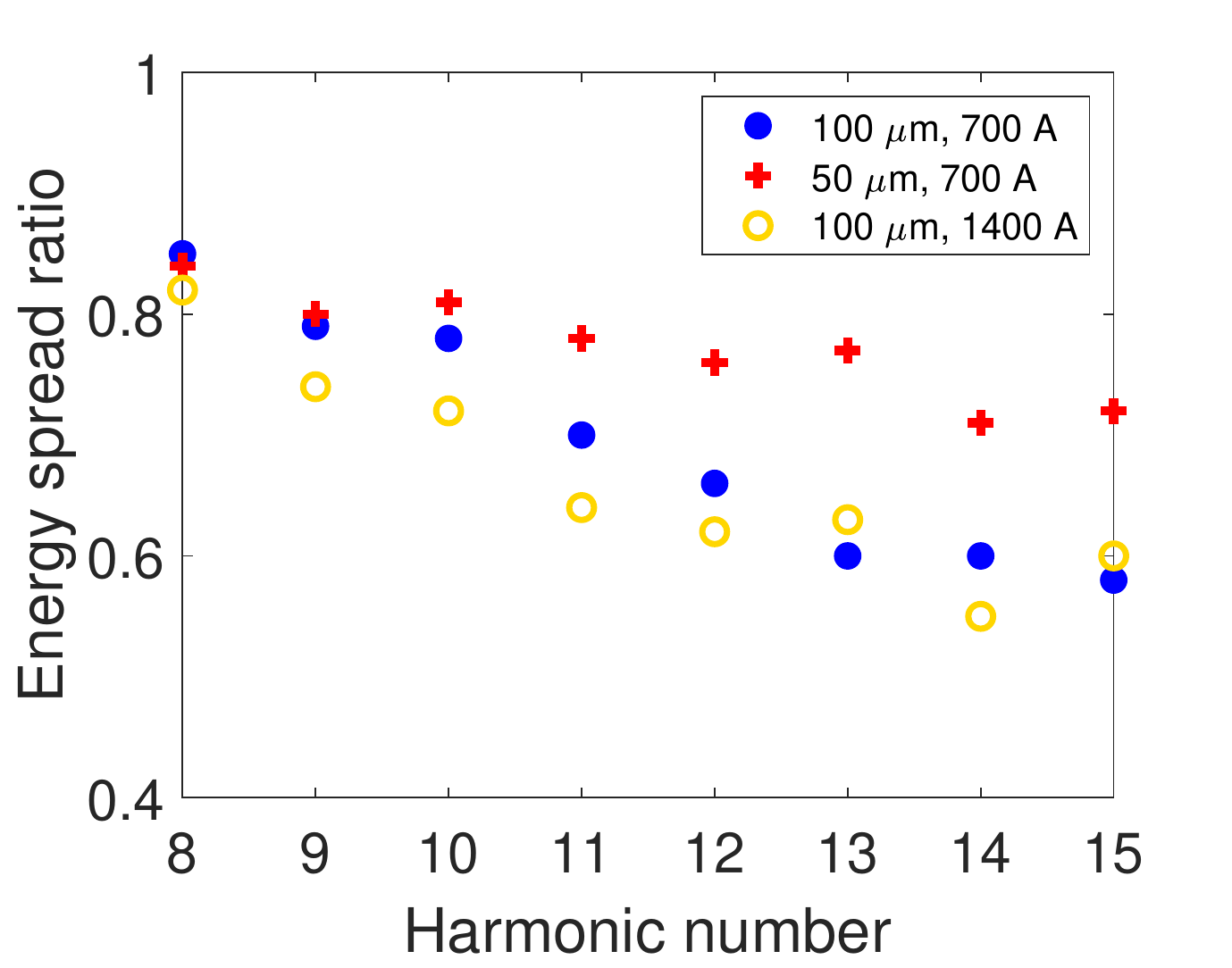}% Here is how to import EPS art
	\caption{\label{fig:4}The energy spread ratio of the standard HGHG and self-modulation HGHG in different cases. The blue dot, red cross, and yellow circle correspond to the nominal case, the second case of beam size of 50 $\mu$m and peak current of 700 A, and the third case of 100 $\mu$m and 1400 A, respectively.}
\end{figure}

The seed laser power ratio represents the ratio of the seed laser power of standard HGHG to that of the self-modulation HGHG, as shown in Fig.~\ref{fig:3}. One can find that the seed laser ratio is a function of the harmonic number. In the nominal case, the seed laser power for the 8th harmonic of the seed laser can be reduced by a factor of 237 and decreases as the number of harmonics increases. In comparison, the beam size of the second case is reduced by half, and others remain constant. In contrast, the peak current of the third case is doubled, as mentioned in Sec~\ref{sec:sec2}. Figure~\ref{fig:3} clearly shows that the beam size and peak current can further reduce the seed laser requirement and enhance the performance of the self-modulation scheme, resulting in a reduction of the seed laser power by up to nearly three orders of magnitude. More importantly, in the nominal case, the pierce parameter of the self-modulator that resonates at the fundamental wavelength of the seed laser is calculated to $\rho = 4.6 \times 10^{-3}$, corresponding to the 1D gain length of 0.80 m and the modulator scaling factor $\eta$ = 2. Similarly, the 1D gain length of the second and third cases is calculated to be 0.51 m and 0.63 m corresponding to the modulator scaling factor of 3.16 and 2.52, respectively. According to Eq. (\ref{eq:6}), the 4-fold scaling curve obtained by the nominal case is shown in Fig.~\ref{fig:3}. Besides, the energy spread radio represents the ratio of the energy spread of standard HGHG to that of the self-modulation HGHG at the entrance of radiators, as shown in Fig.~\ref{fig:4}. Owing to the CHG generation in the self-modulator, the energy spread of self-modulation HGHG is larger than that of standard HGHG. Figure~\ref{fig:4} illustrates that the energy spread of the second case by reducing the beam size is smaller than that of the third case by increasing the peak current. Since the three-dimensional (3D) effect and the additional energy modulation introduced by the CHG radiation spot size in the self-modulator are neglected in the roughly estimated Eq. (\ref{eq:6}), the second case shows that the ratio is poorly conformed at higher harmonics; the third case shows that the ratio is better conformed, but with a slight increase on higher harmonics over 12th. Therefore, the effect of beam size $\sigma_x$ and peak current $I$ can still be estimated roughly by Eq. (\ref{eq:6}), and modulator scaling factor $\eta$ is an intermediate variable during the design and optimization of self-modulation HGHG.
\begin{figure}
\centering
	\includegraphics[width=0.24\textwidth]{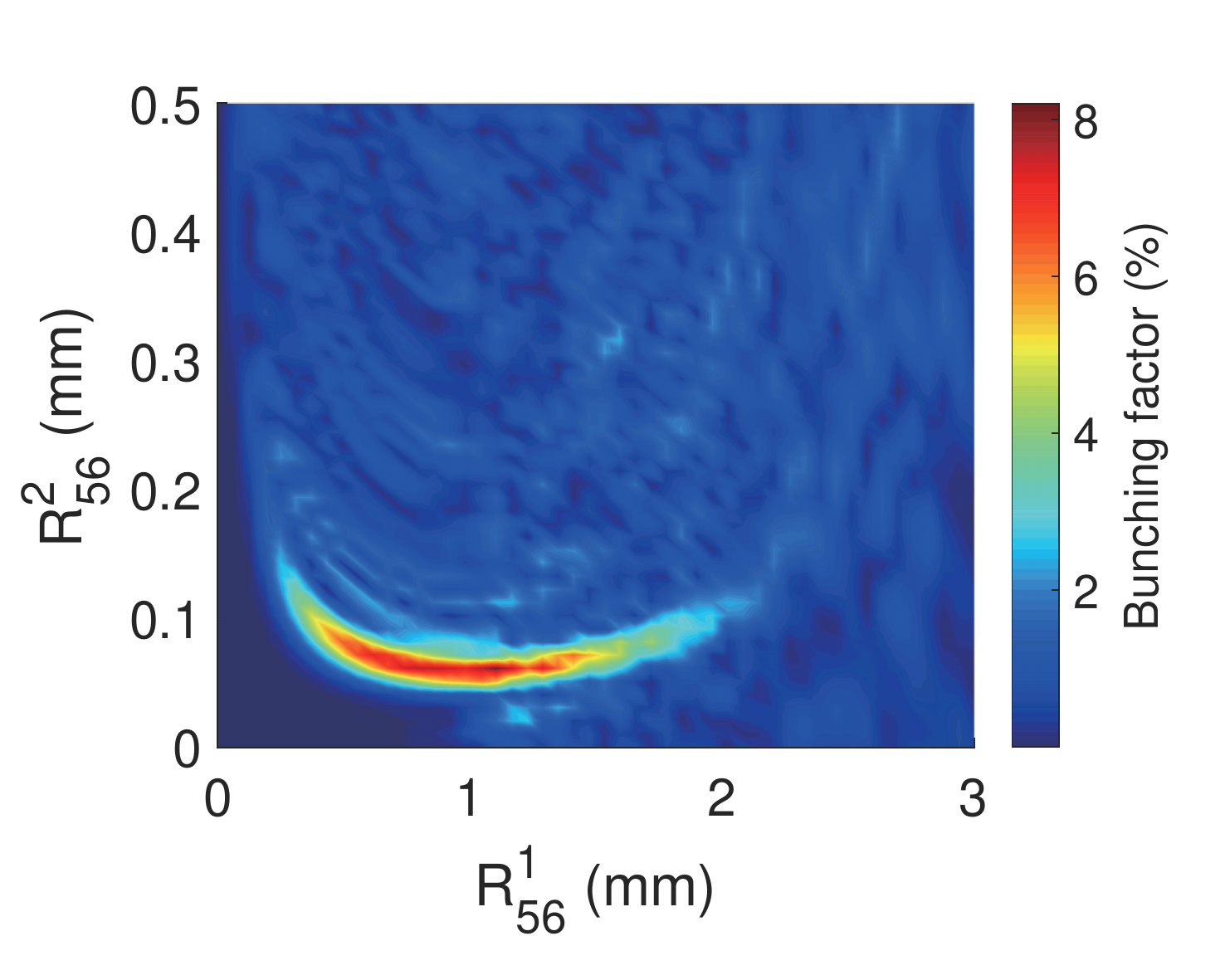}
	\includegraphics[width=0.23\textwidth]{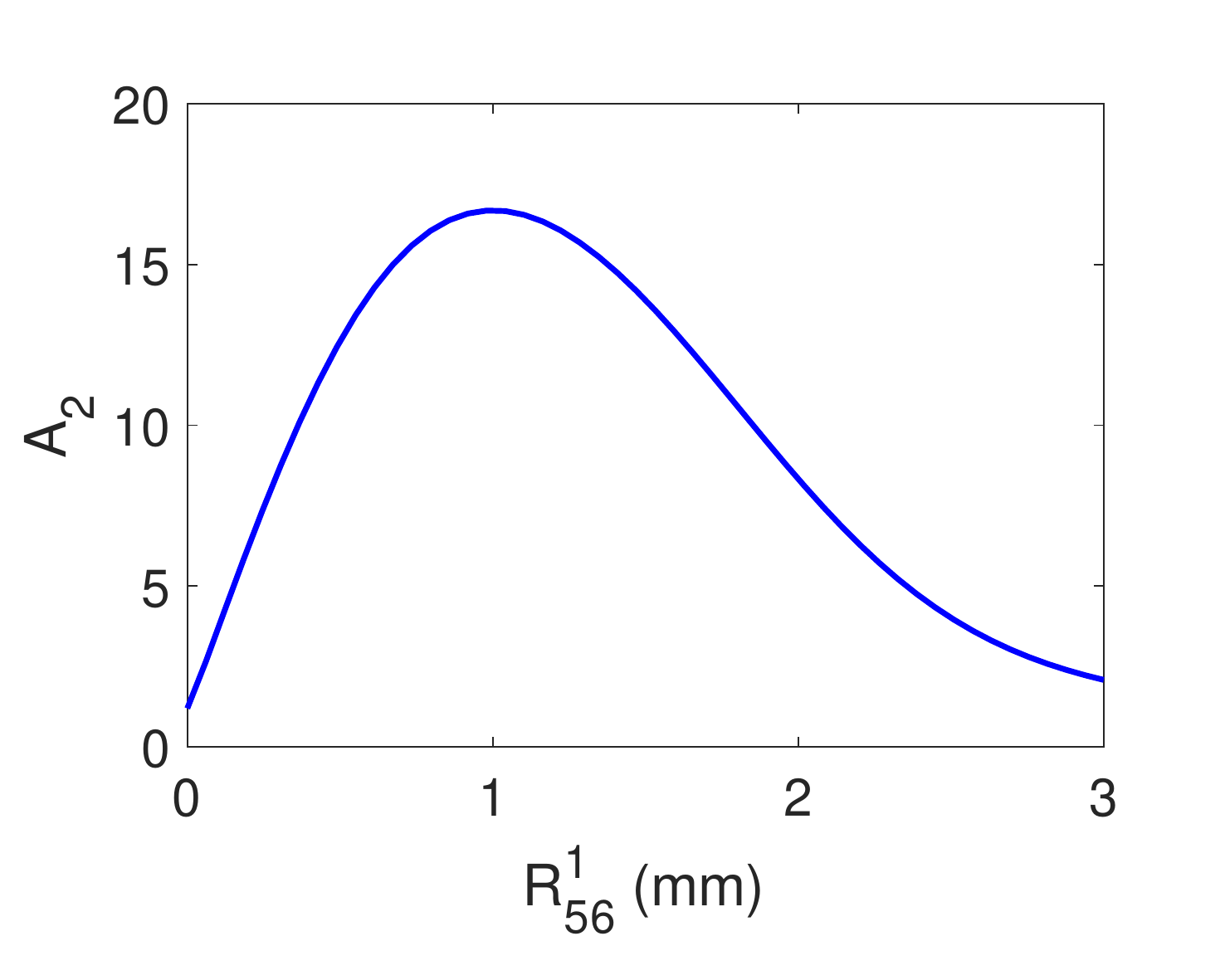}\\
	\includegraphics[width=0.24\textwidth]{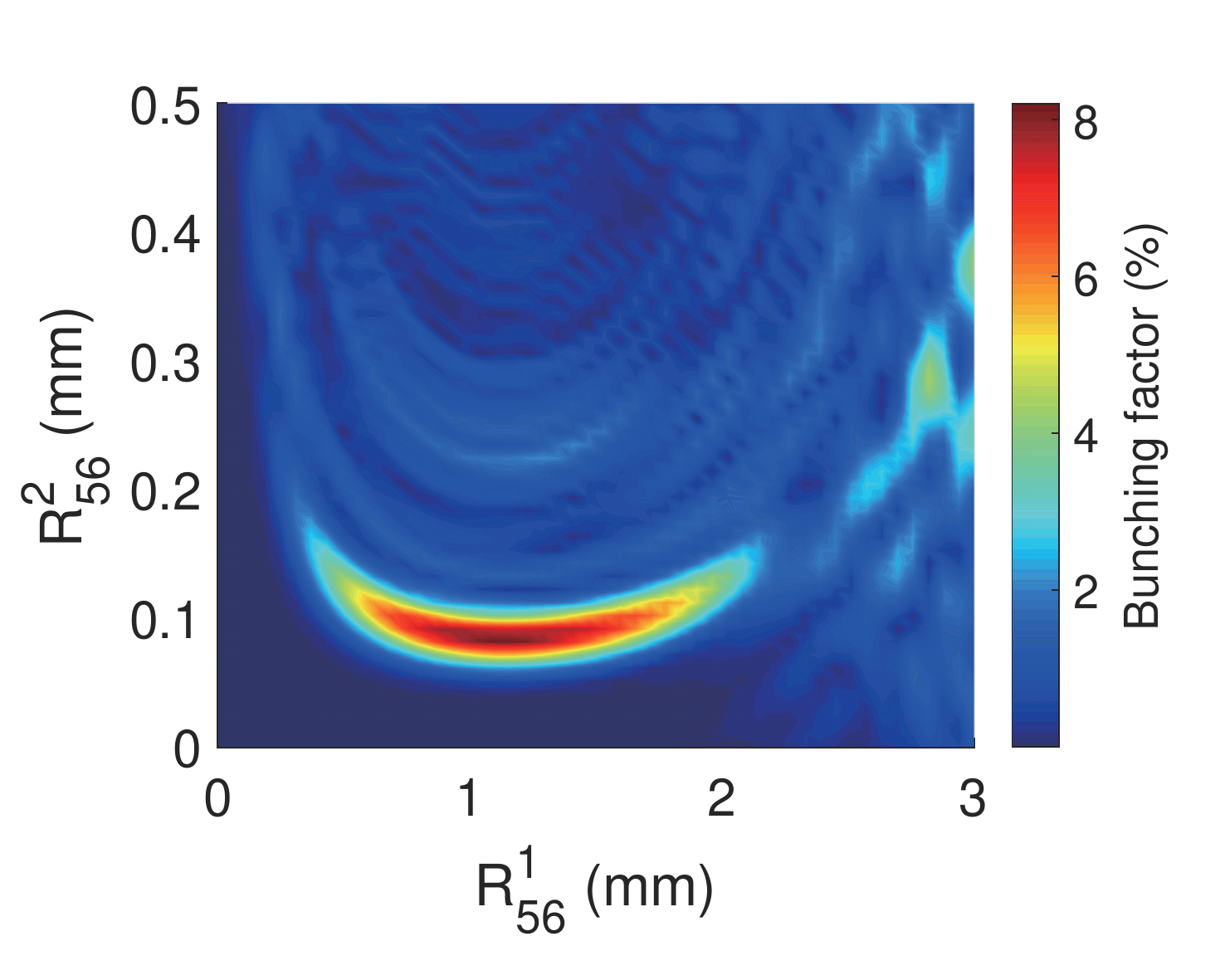}
	\includegraphics[width=0.23\textwidth]{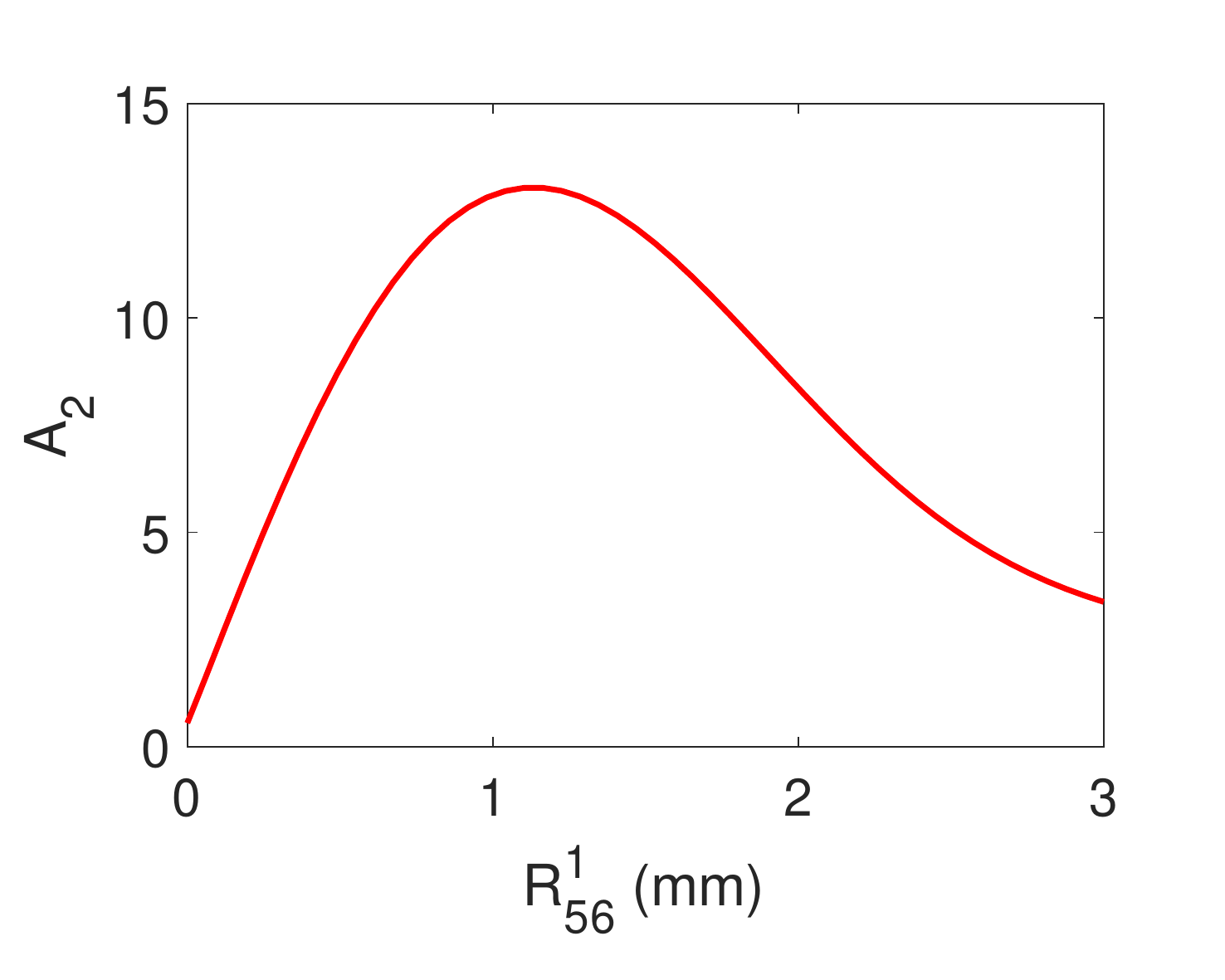}
	\caption{\label{fig:5}Optimization for the $R_{56}$ of two chicanes by GENESIS simulations to obtain the 13th harmonic bunching factor of 8\% and corresponding energy modulation amplitude $A_2$ of the entrance of the radiator in different beam sizes. The top and the bottom correspond to the beam size of 100 $\mu$m and 50 $\mu$m, respectively.} 
\end{figure}

\begin{figure*}[htb]
\hspace{-12pt}
	\includegraphics[width=0.25\textwidth]{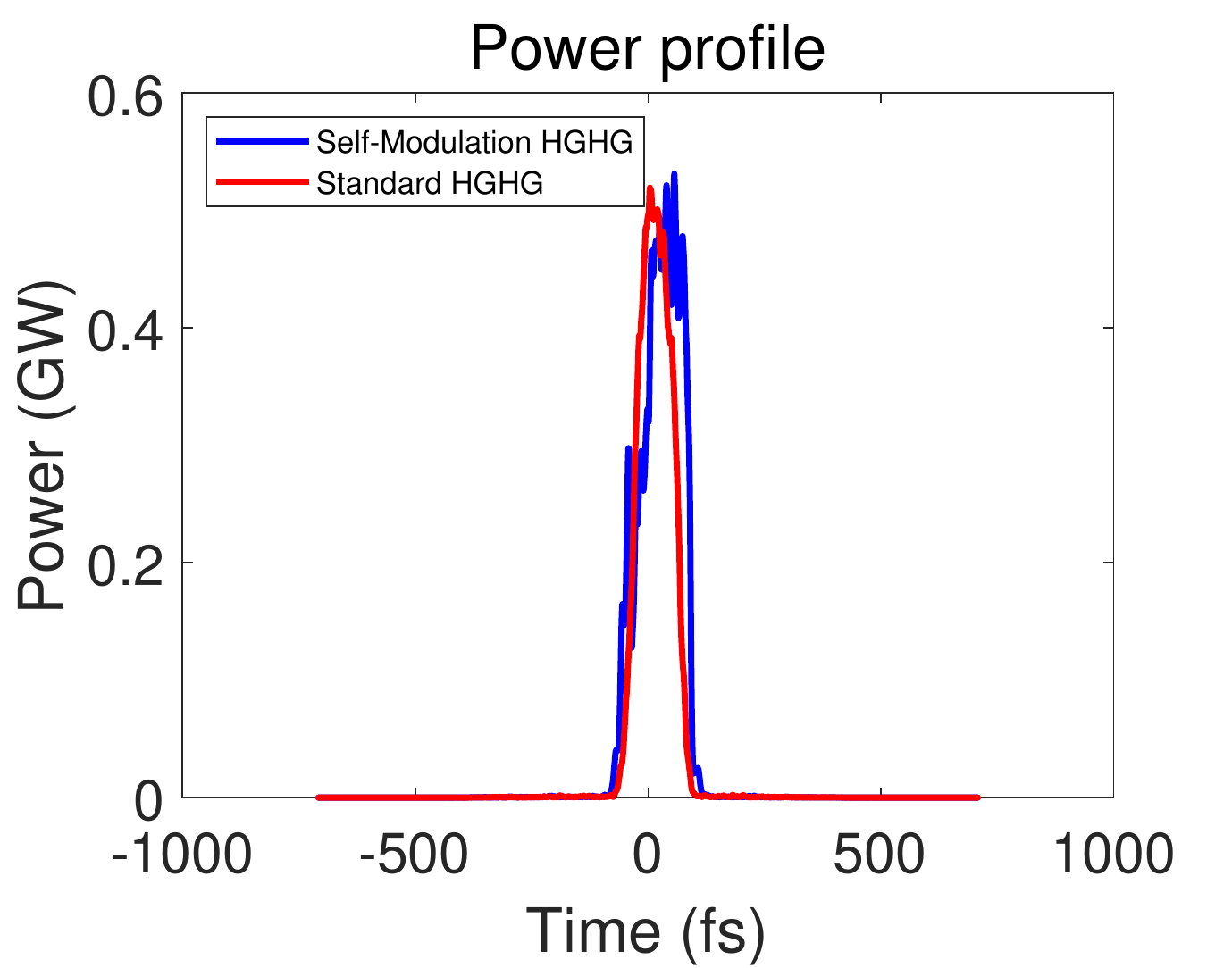}
	\includegraphics[width=0.25\textwidth]{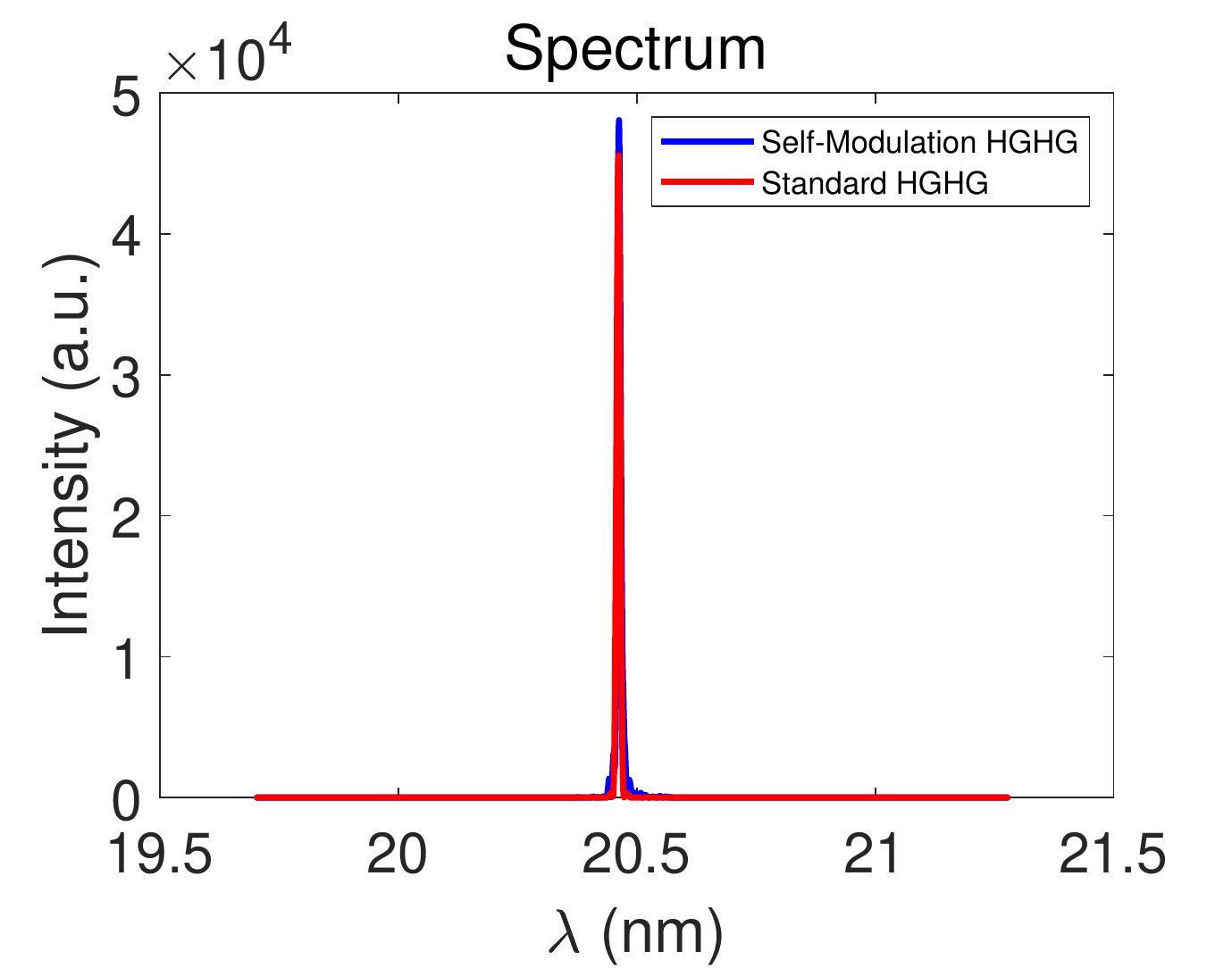}
	\includegraphics[width=0.25\textwidth]{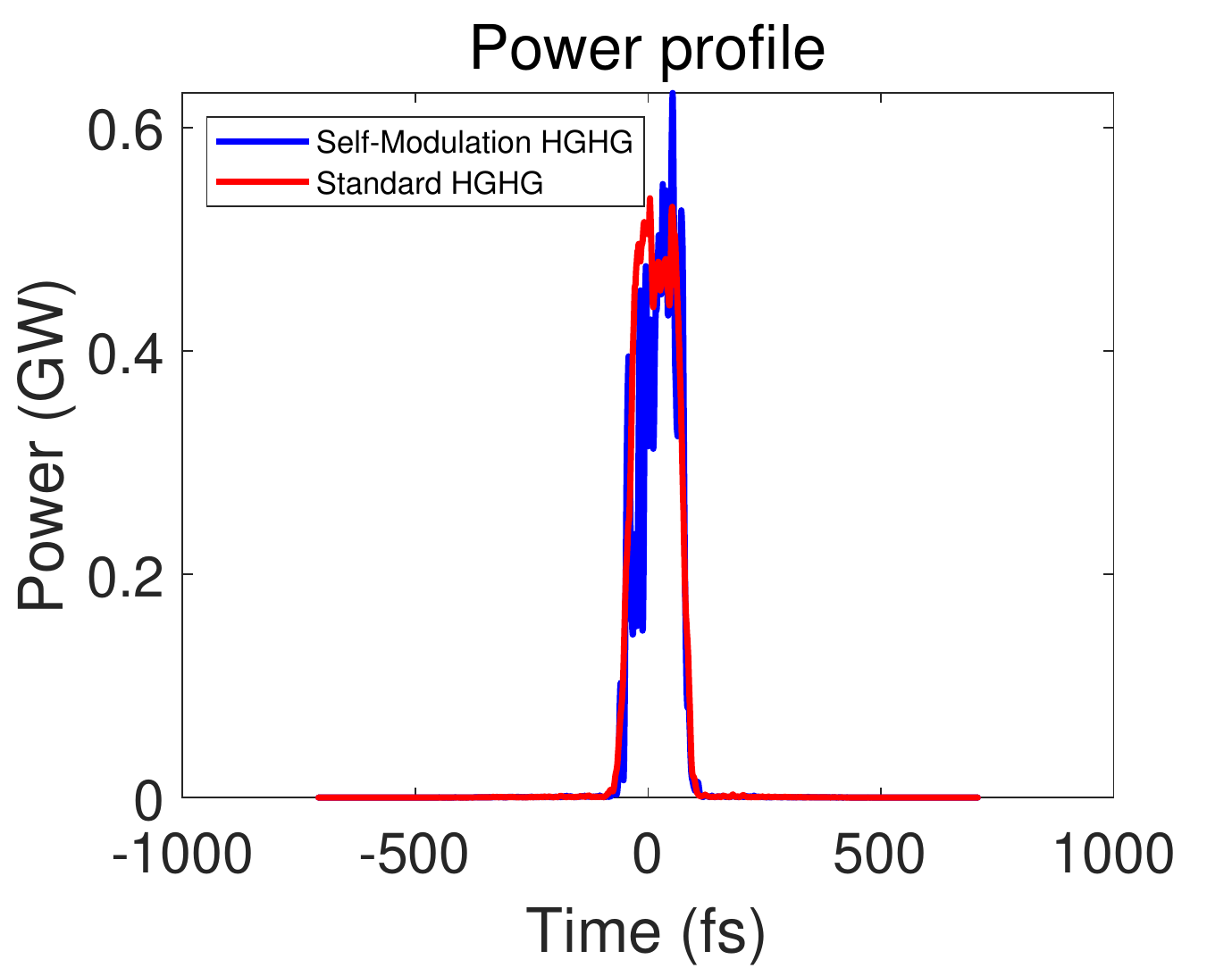}
	\includegraphics[width=0.25\textwidth]{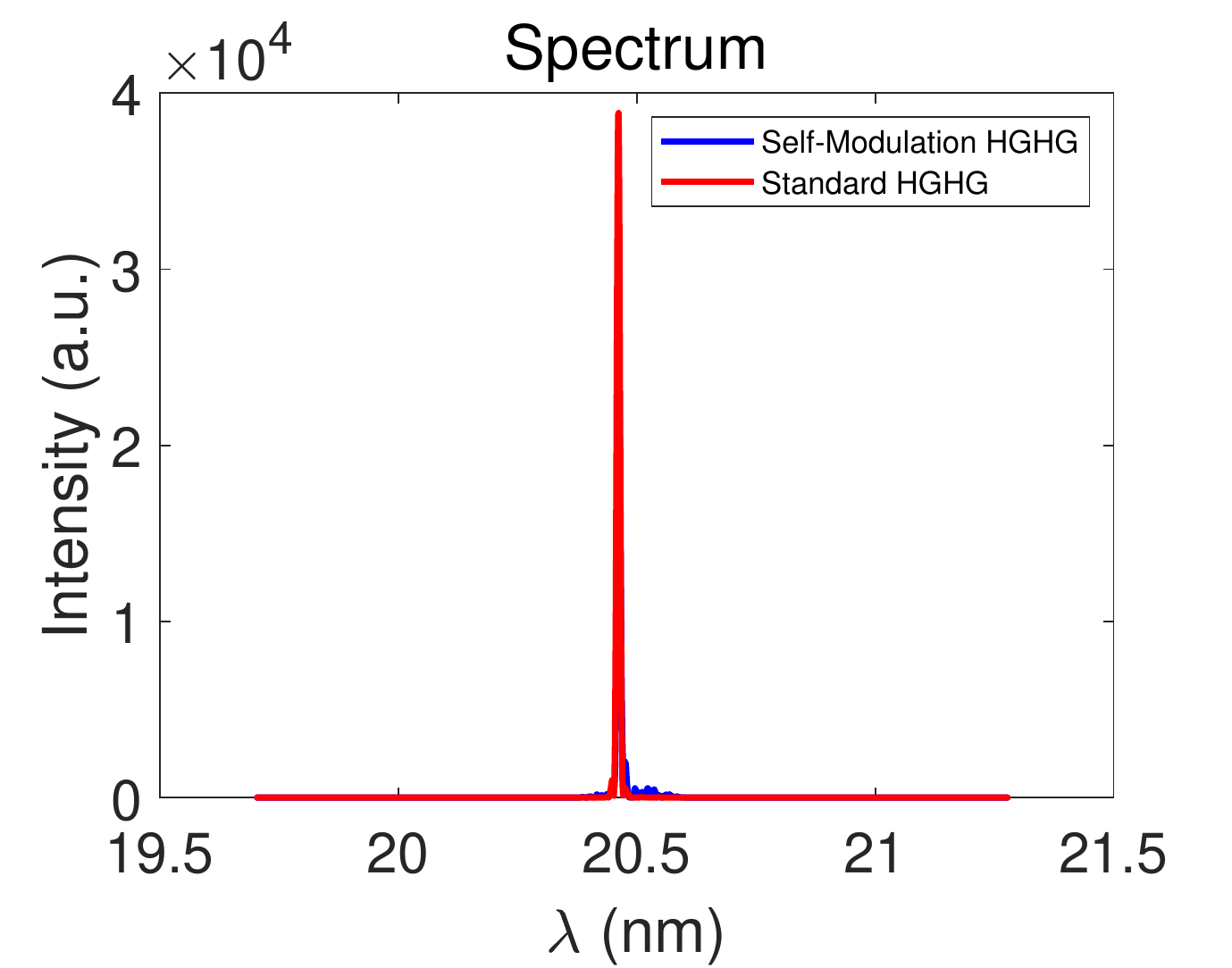}
	\caption{\label{fig:6}Comparison of the FEL performance between self-modulation HGHG and standard HGHG at the 13th harmonic of the seed laser with the wavelength of 266 nm in different beam sizes. The left and the right correspond to the beam size of 100 $\mu$m and 50 $\mu$m, respectively.} 
\end{figure*}

\begin{table*}[htb]
	\caption{\label{tab:table4}The main parameters of output FEL at the 13th harmonic of the seed laser in standard HGHG and self-modulatio HGHG with different beam sizes.}
    \begin{ruledtabular}
        \begin{tabular}{ccccc}
        &\multicolumn{2}{c}{Standard HGHG}&\multicolumn{2}{c}{Self-Modulation HGHG} \\
        Beam size ($\mu$m)           & 100 & 50 & 100       & 50                                \\ \hline
        Peak power@12m (GW)     & 0.519     & 0.537          & 0.531                & 0.631     \\
        Pulse length (fs)       & 93.100& 113.460         & 133.000                           & 121.473      \\
        Central wavelength (nm) & 20.460  & 20.461    & 20.463   & 20.468      \\
        Bandwidth (\%)          & 0.167    & 0.155   & 0.188     & 0.281 \\
        Pulse energy ($\mu$J)        & 46.772  & 59.729  & 55.742   & 51.940                           
        \end{tabular}
    \end{ruledtabular}
\end{table*}

To evaluate the reduction of seed laser power requirement, we carry out an example analysis of self-modulation HGHG at the 13th harmonic of the seed laser. Initially, in different beam sizes of 100 $\mu$m and 50 $\mu$m, we introduce an optimal energy modulation in the first modulator by a 266-nm seed laser with peak power of 0.62 MW and 0.085 MW to the initial energy modulation $A_1$ of 1.2 and 0.6, respectively. The beam size is reduced by half, corresponding to a 7.3-fold reduction of the seed laser power. Secondly, the two-dimensional scanning of the $R_{56}$ of dispersive chicanes for obtaining the 13th harmonic bunching factor of 8\% in different beam sizes is shown in Fig.~\ref{fig:5}. In the case of the beam size of 100 $\mu$m, the optimal $R_{56}$ values of the first and second chicanes are 1.04 mm and 0.06 mm, respectively. Similarly, the optimal $R_{56}$ values of the first and second chicane in the other case are 0.92 mm and 0.09 mm, respectively. It is found that $R_{56}^2$ is more sensitive than $R_{56}^1$. Thirdly, with the energy spread of the entrance of radiators, the energy modulation amplitude $A_2$ can be roughly estimated by $\sqrt{2 [(\sigma_{\gamma^{'}} / \sigma_{\gamma})^2 - 1]}$. In the case of the beam size of 100 $\mu$m, the energy modulation is about 850 keV corresponding to $A_2$ = 16.7, while a lower energy modulation introduced in the case of the beam size of 50 $\mu$m is only about 650 keV corresponding to $A_2$ = 13.0. The magnification efficiency of energy modulation for $A_2/A_1$ is about 13.9 and 21.7 in different beam sizes, respectively. Thus, the beam size reduction significantly can relax the seed laser power requirement and increases the bunching factor's efficiency. Finally, the 3D time-dependence simulation are carried out, and then the output FEL performance of the 13th harmonic of the seed laser at 12 m along the radiators is shown in Fig.~\ref{fig:6}, as summarized in Table~\ref{tab:table4}. In the comparison between the self-modulation HGHG and standard HGHG with different beam sizes, the seed laser required for the self-modulation HGHG is only a few tens of kW, and the shot noise in the first modulator is about 43 W, which can be estimated by \cite{saldin1999physics,Marinelli2008}
\begin{equation}
P_{\mathrm {noise}}=\frac{3^{3 / 4} 4 \pi \rho^{2} P_{\mathrm {beam }}}{N_{\lambda} \sqrt{\pi} \eta},
\end{equation}
where $P_{\mathrm{beam}}$ is the electron beam power, and $N_{\lambda}$ is the number of electrons per seed laser wavelength. The signal-to-noise ratio is significantly lower than that of the standard HGHG with the introduction of a few tens of MW seed lasers. Compared to the standard HGHG, the pulse length or bandwidth slightly increases in the self-modulation HGHG. However, the degradation of longitudinal coherence is not obvious and seed laser power is easily reduced by a factor of around 932.

\section{\label{sec:sec4}Towards short wavelength}
Typically, short-wavelength seeded FELs are limited by two main factors: seed laser performance, such as wavelength $\lambda_s$ and peak power, and harmonic up-conversion efficiency for high-gain FELs. As described in Sec.~\ref{sec:sec3}, the self-modulation scheme can significantly reduce the seed laser requirement. To investigate the potential of the self-modulation scheme to generate ultra-high harmonics, the self-modulation HGHG is further discussed.
\begin{figure}
	\includegraphics[width=0.24\textwidth]{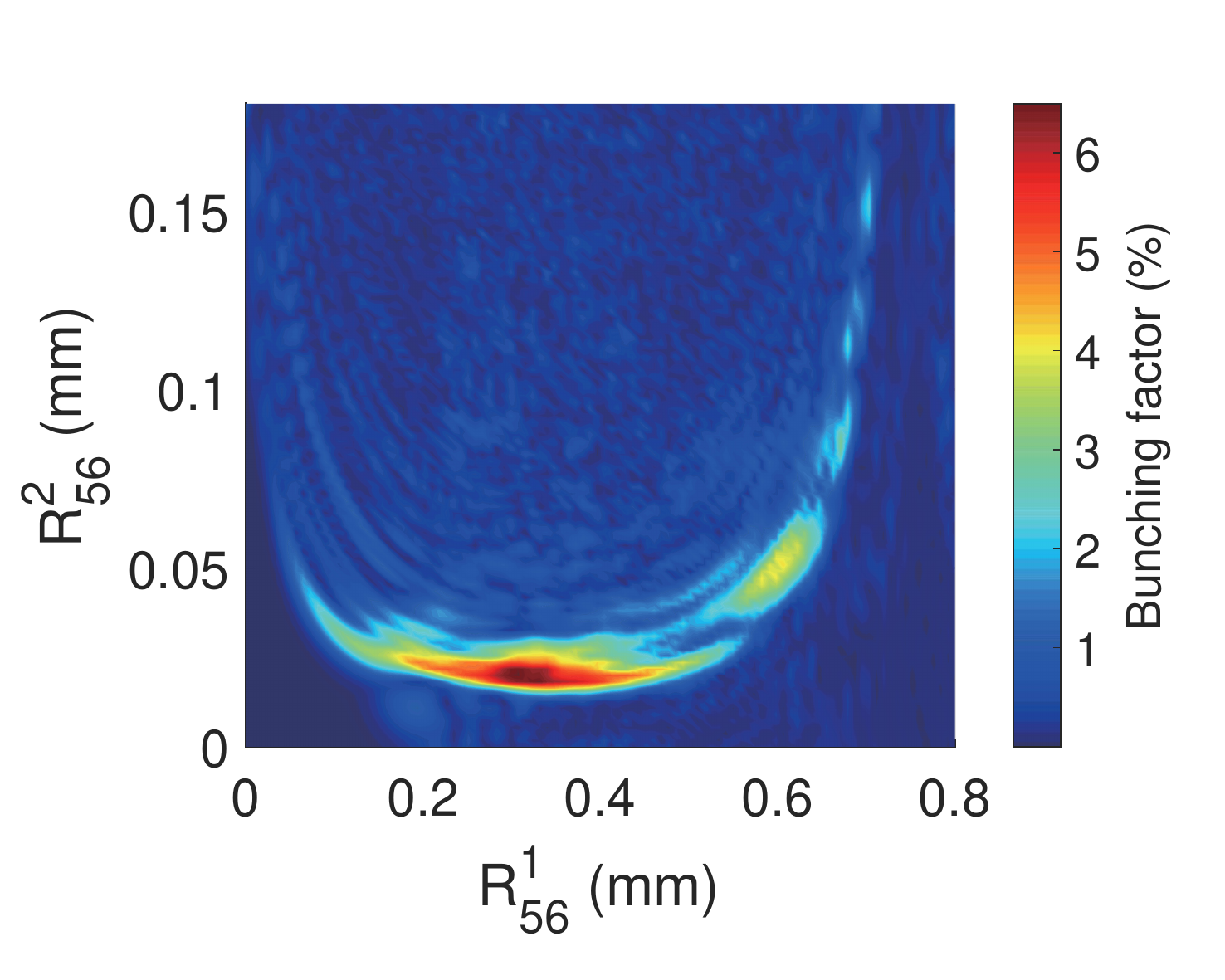}
	\includegraphics[width=0.23\textwidth]{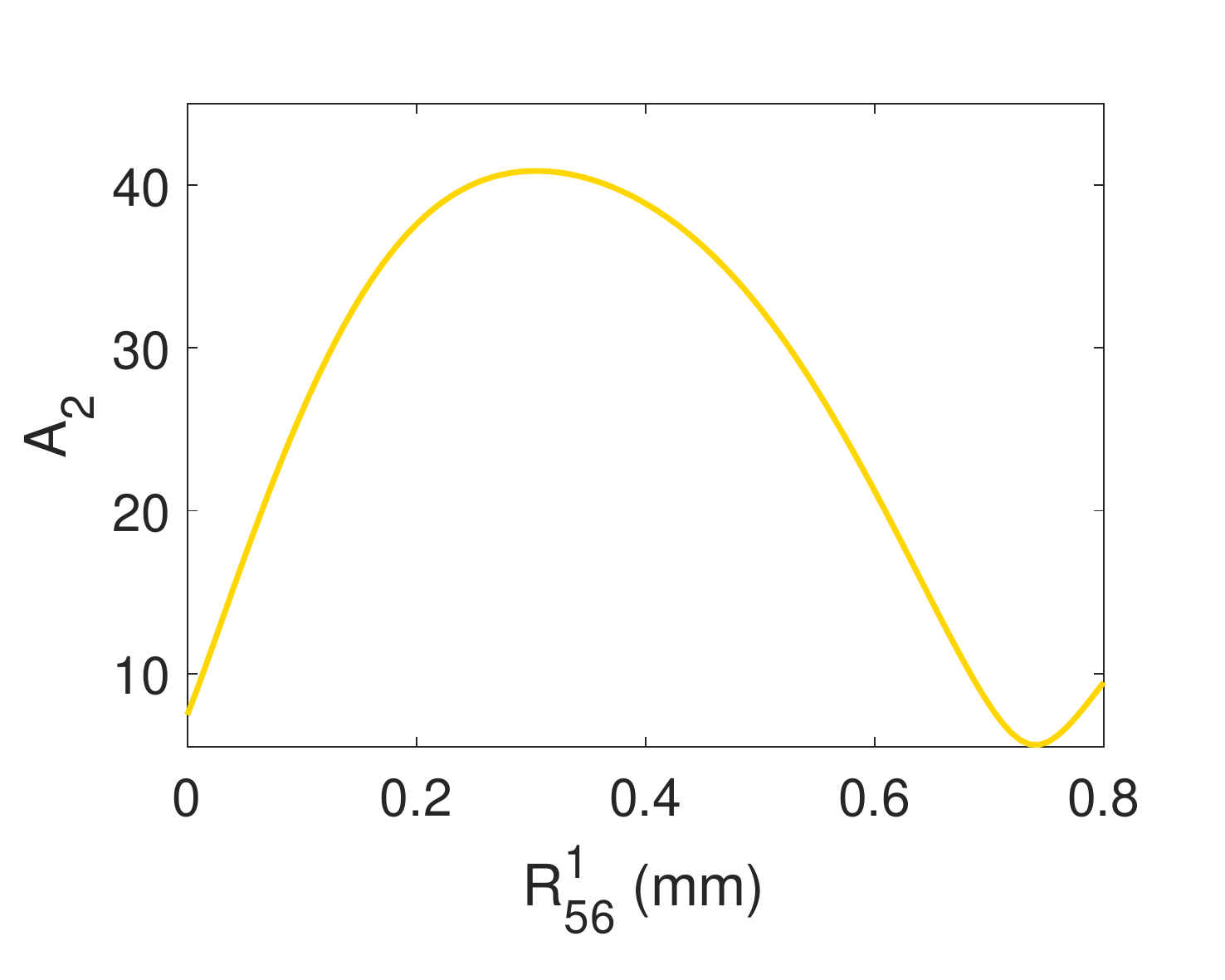}\\
	\includegraphics[width=0.24\textwidth]{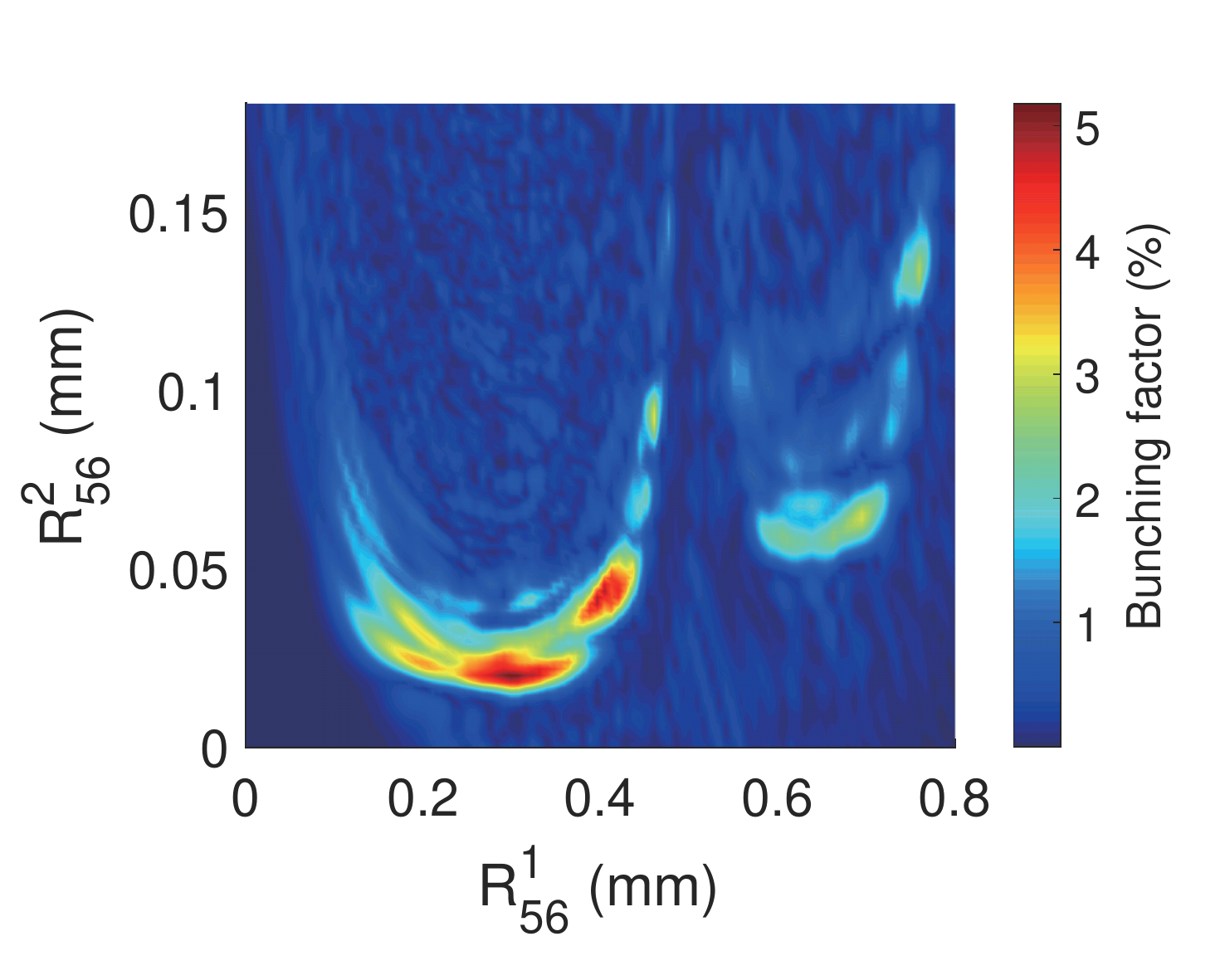}
	\includegraphics[width=0.23\textwidth]{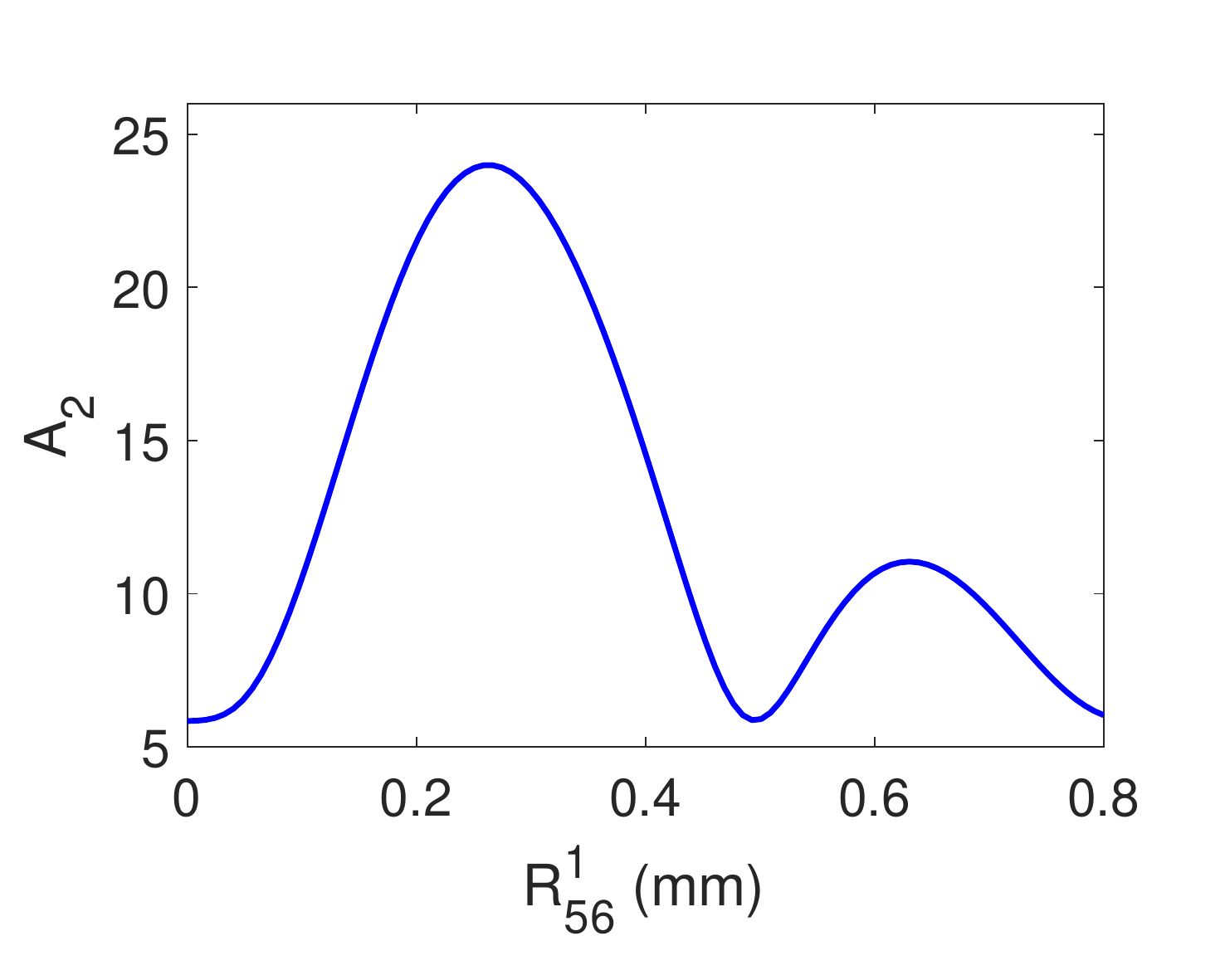}\\
	\includegraphics[width=0.24\textwidth]{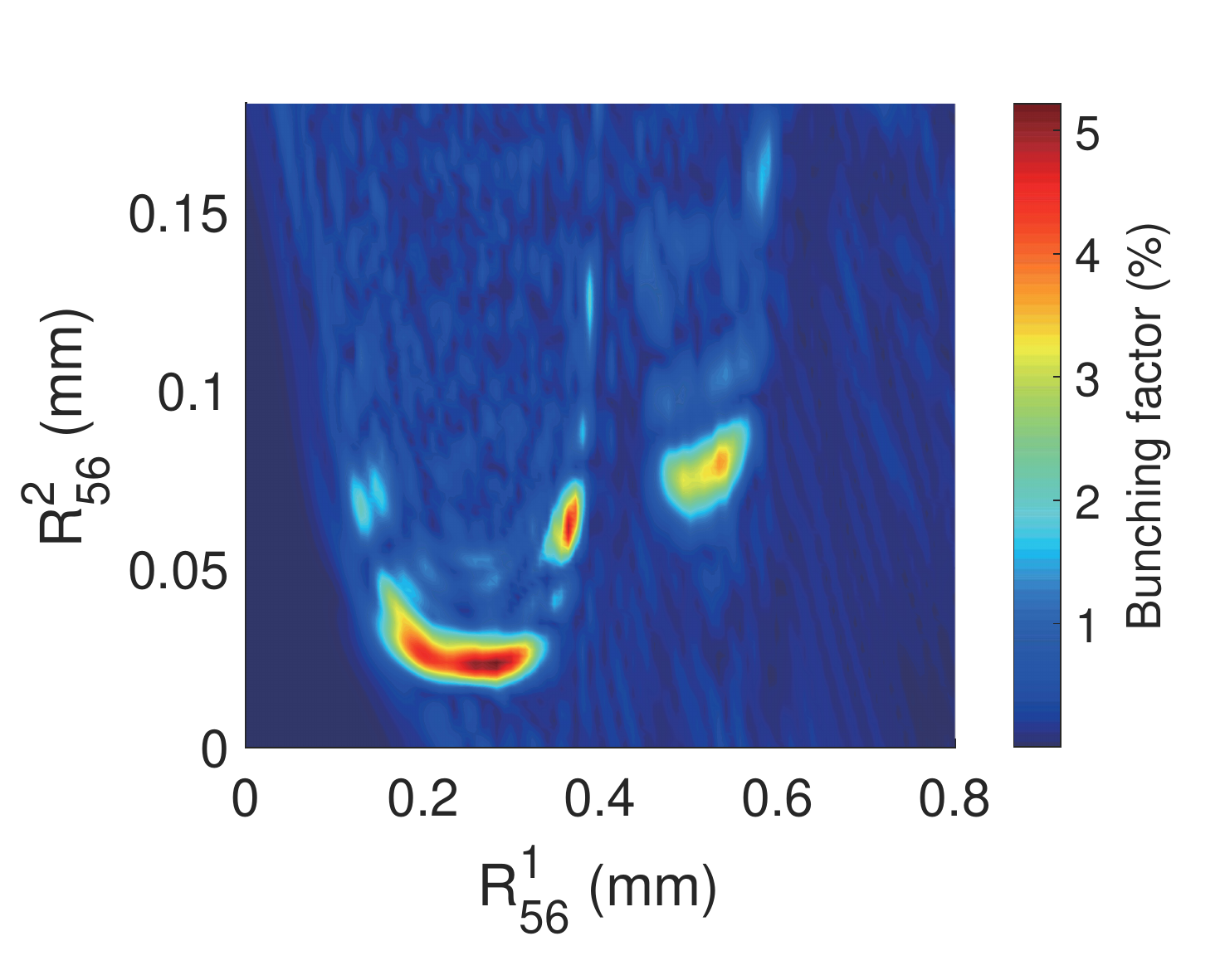}
	\includegraphics[width=0.23\textwidth]{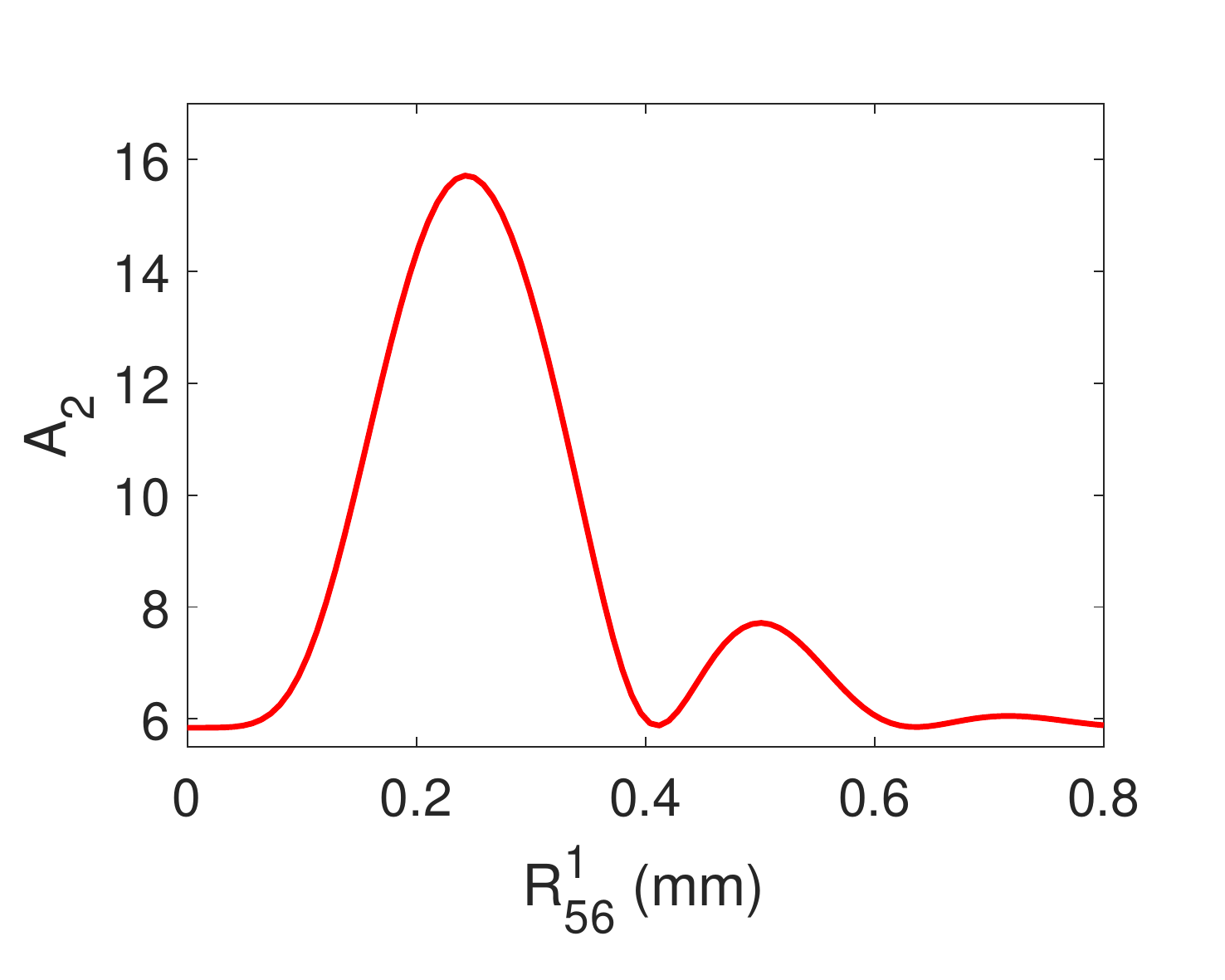}
	\caption{\label{fig:7}Optimization for the $R_{56}$ of two chicanes towards the 30th harmonic of the seed laser. The self-modulator resonates at the fundamental wavelength (top), the second (middle), and the third (bottom) harmonics of the seed laser, respectively.}
\end{figure}

\begin{figure}[t]
    \hspace{-12pt}
	\includegraphics[width=0.25\textwidth]{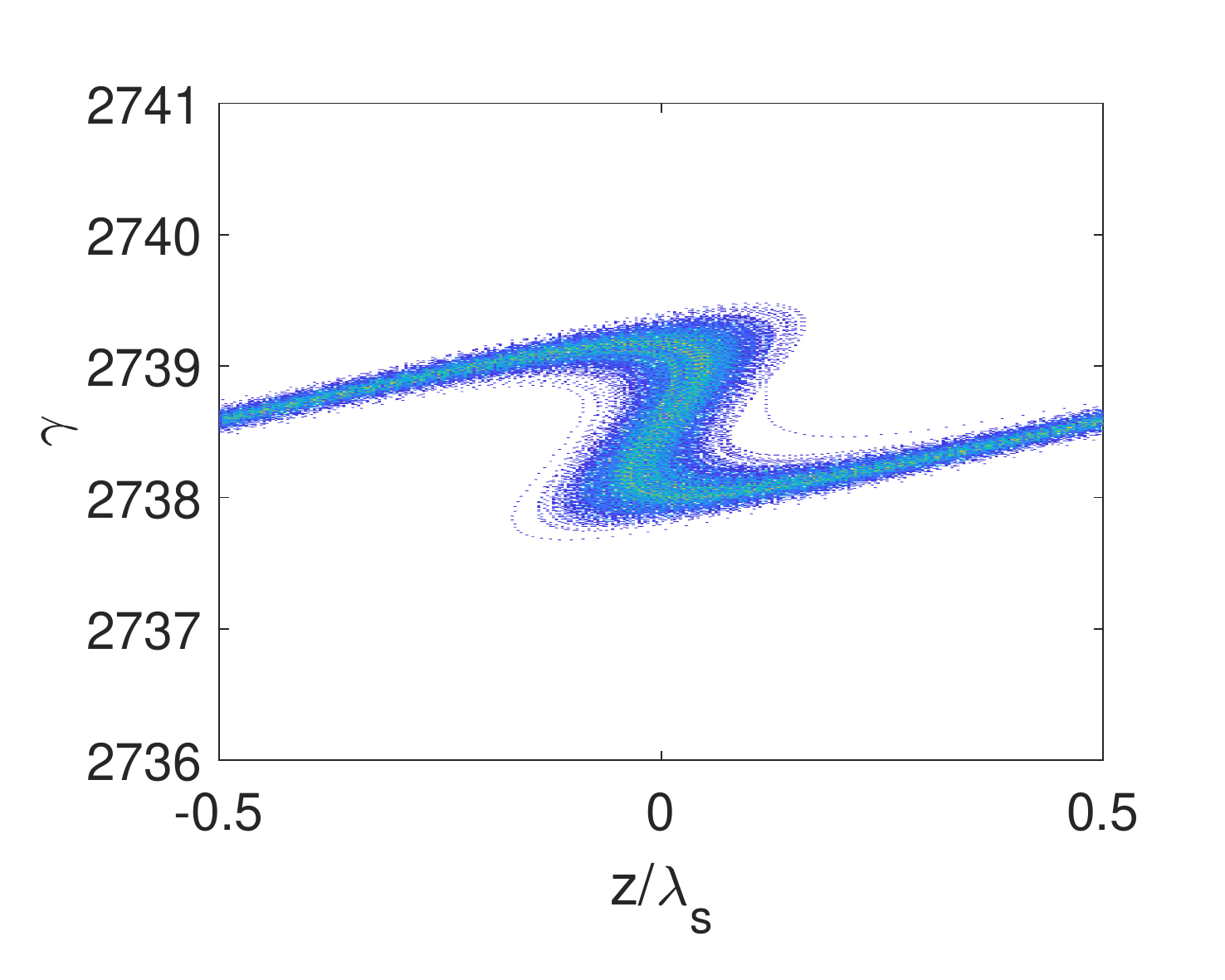}% Here is how to import EPS art
	\includegraphics[width=0.25\textwidth]{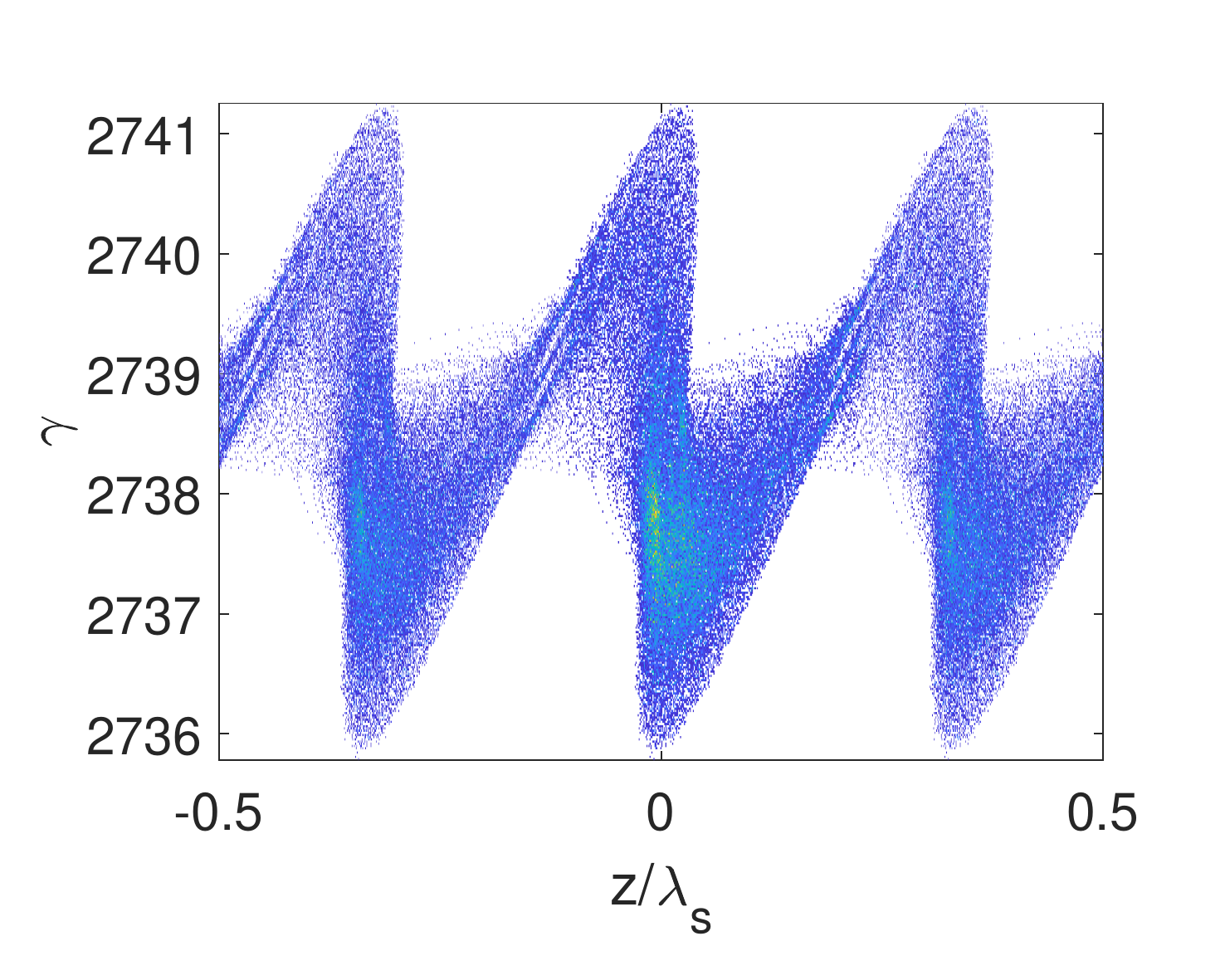}% Here is how to import EPS art
	\caption{\label{fig:8}The longitudinal phase space of the electron beam in one seed laser wavelength $\lambda_{s}$ at the entrance of the self-modulator (left) and radiator (right), where the self-modulator is tuned at the third harmonic of the seed laser.}
\end{figure}

\begin{figure*}
	\includegraphics[width=0.3\textwidth]{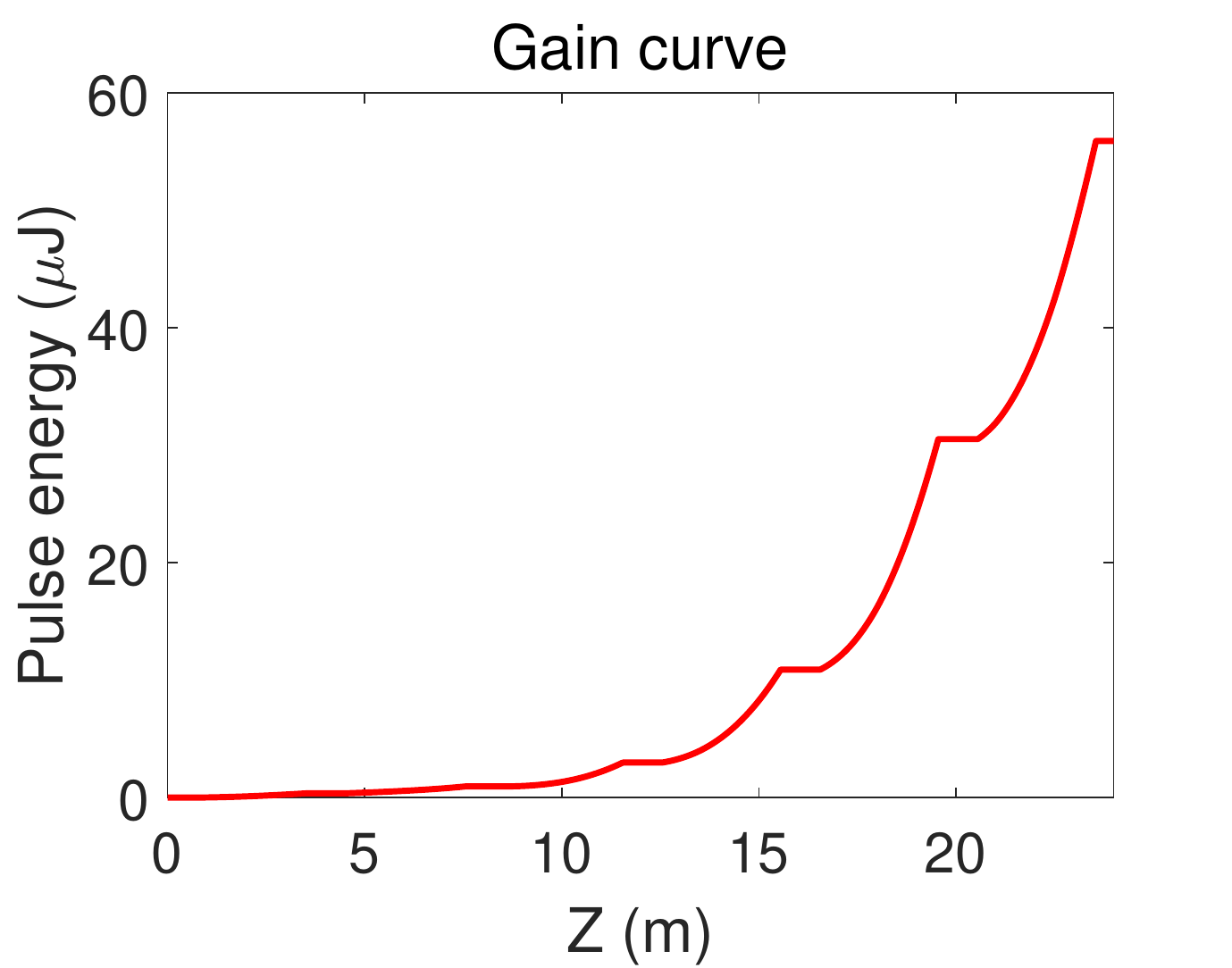}
	\quad
	\includegraphics[width=0.3\textwidth]{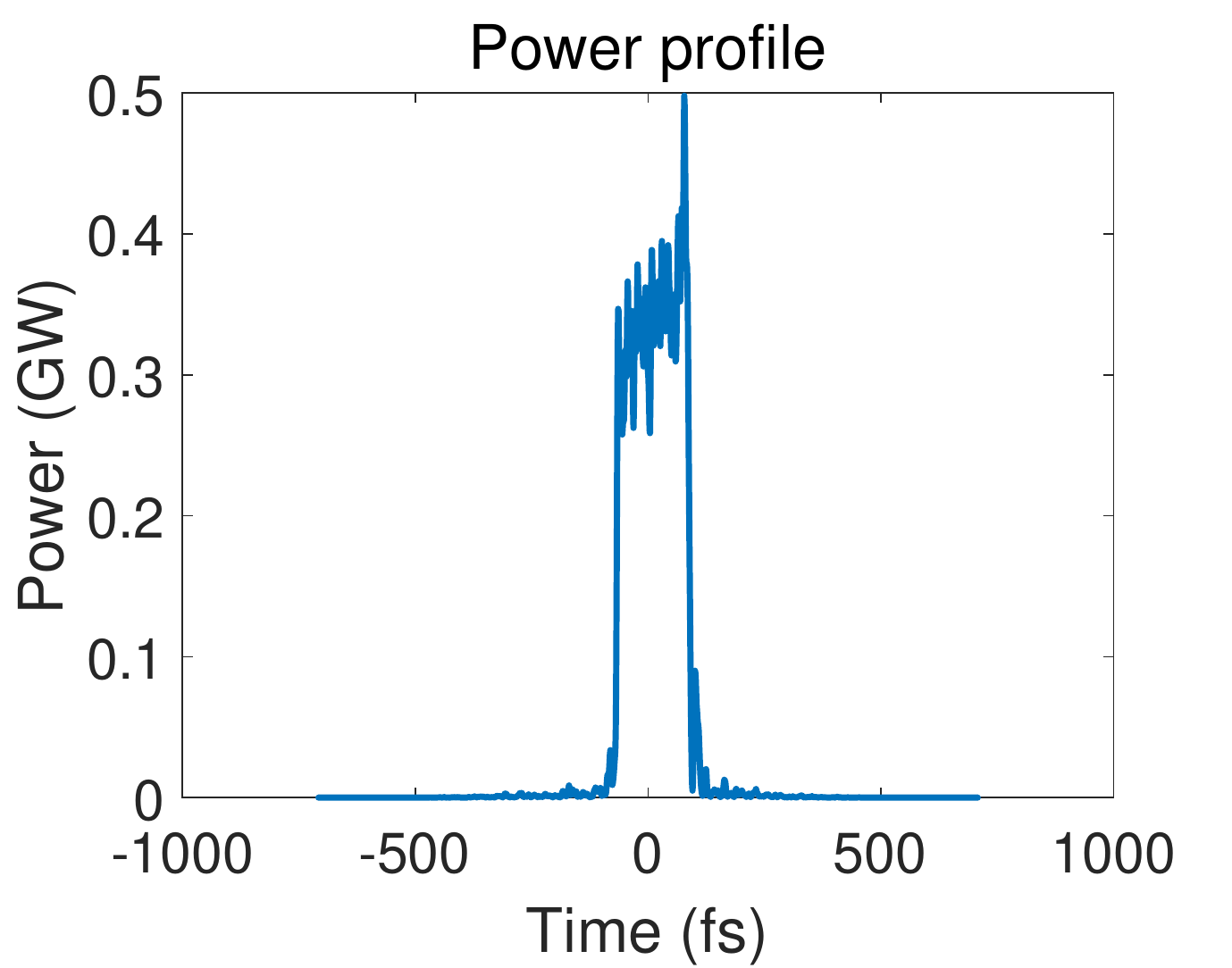}
	\quad
	\includegraphics[width=0.3\textwidth]{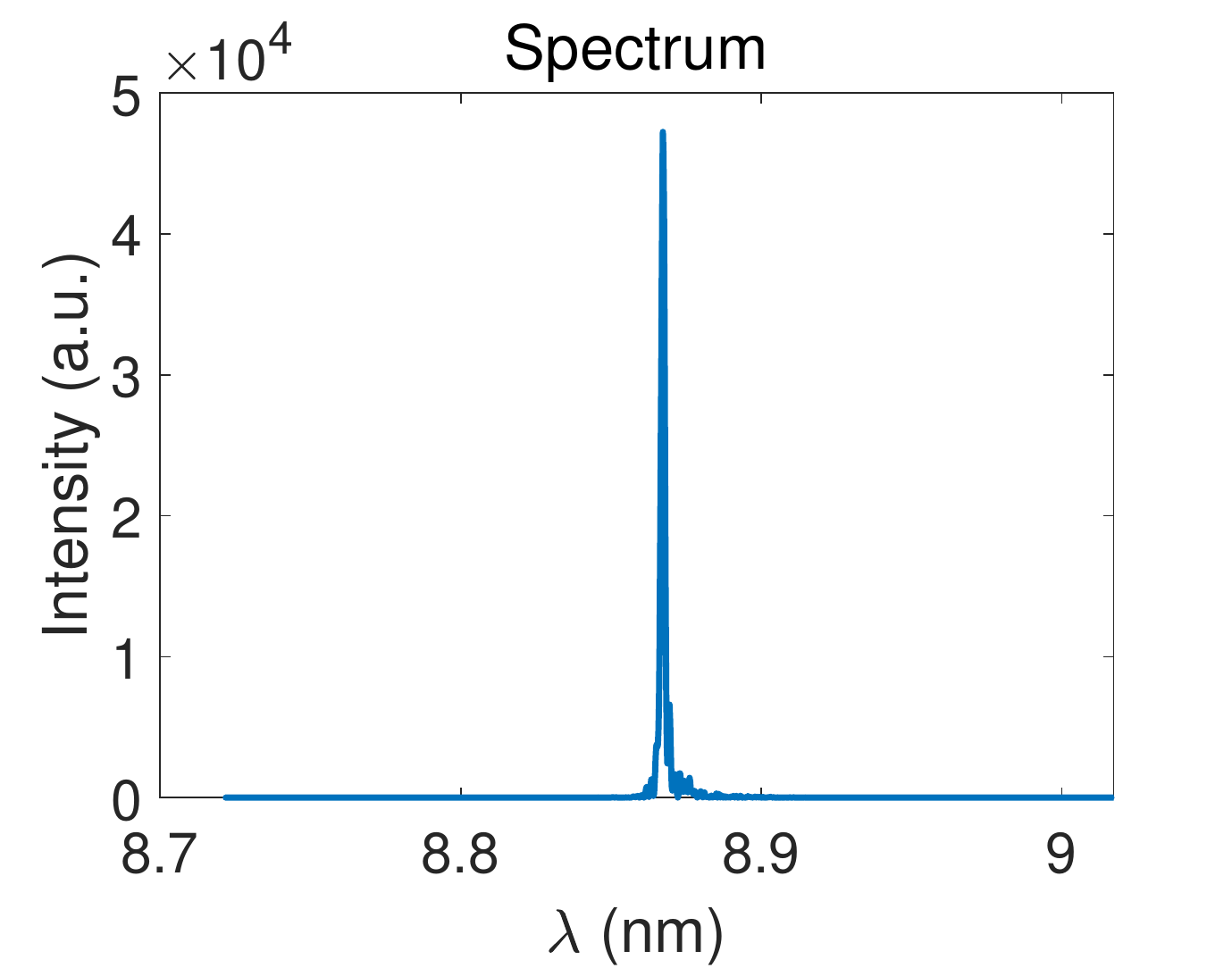}
	\caption{\label{fig:9}The output FEL performance at the 30th harmonic of the seed laser in the third-harmonic self-modulation. The left corresponds to the 8.87 nm radiation gain curve in the radiator. The middle and the right correspond to the power profile and the spectrum after six radiator modules, respectively.}
\end{figure*}

With the parameters of the nominal case listed in Table~\ref{tab:table1}, the optimization of the self-modulation scheme at SXFEL-UF is presented in Fig.~\ref{fig:7} for a short-wavelength FEL. To enhance the coherent energy modulation, we increase the self-modulator length from 1.6 m to 2 m in this section, corresponding to $\eta$ = 2.5. To obtain an adequate bunching factor in various harmonics and improve the signal-to-noise ratio, the initial energy modulation amplitude $A_1$ is significantly increased to 6, corresponding to the seed laser peak power of 20 MW. As shown in Fig.~\ref{fig:7}, the bunching factor larger than 5\% is obtained when the self-modulator resonates at the fundamental, second, and third harmonics of the seed laser. The optimal dispersion strength $R_{56}^1$ is considered almost unchanged, and the harmonic bunching is more sensitive to the second dispersion strength $R_{56}^2$. The large energy modulation amplitude $A_2$ = 40 is obtained in the case of self-modulator resonated at the fundamental wavelength. However, the energy spread is too large to generate high-gain FEL pulses in this case. Besides, at least 5\% of the bunching factor can be obtained when the self-modulator resonates at the second and third harmonic, corresponding to the maximum energy modulation amplitude of 24 and 15, respectively. Figure~\ref{fig:7} also indicates that the coherent energy modulation amplitude $A_2$ decreases significantly as the tuned harmonic number rises at the self-modulator. Therefore, the self-modulator can resonate at a higher harmonic, generating an appropriate energy modulation with a lower energy spread.

As analyzed in Sec.~\ref{sec:sec2}, the self-modulator can produce a large energy modulation with a weak initial energy modulation, where the self-modulator resonates at the fundamental wavelength. Moreover, the maximum energy modulation in the self-modulation HGHG is significantly larger than that of the standard HGHG when both configurations have the same harmonic bunching factor. Due to the additional energy spread introduced by the self-modulation process, the total energy spread of the self-modulation HGHG is significantly larger than that of the standard HGHG, especially at higher harmonics (see Fig.~\ref{fig:4}). As discussed above, in the third-harmonic self-modulation, the induced energy spread is minimum, while the bunching at the 30th harmonic is still considerable. The 3D time-dependent simulations are carried out to verify the feasibility of the third-harmonic self-modulation. After an optimal $R_{56}$, the energy modulation can be converted into density distribution with the 30th harmonic bunching factor of 5\%. As shown in Fig.~\ref{fig:8}, the longitudinal phase space of the self-modulation HGHG is not ideally sinusoidal energy modulation compared with the standard HGHG. It is worth noting that the degradation of energy spread can be overcome to achieve ultra-higher harmonics in the self-modulation HGHG. Further enhancement of the initial energy modulation with self-modulator resonated at higher harmonics can increase harmonic up-conversion while significantly avoiding the electron beam quality degradation.

Furthermore, the output FEL gain curve, power profile, and spectrum are summarized in Fig.~\ref{fig:9}, using the SXFEL-UF parameters listed in Table~\ref{tab:table2} and \ref{tab:table3}. The saturation pulse energy reaches 54.7 $\mu$J at the end of radiators, whose total length of 24 m. Because of the additional slippage in the self-modulator and the pulse shorten effect in seeded FELs, the final output FEL pulse length of about 153.4 fs (FWHM) is comparable with the seed laser pulse duration of 150 fs (FWHM). There are several spikes in the power profile due to the lower input seed laser power and the shot noise effect. After six radiator modules, the maximum output peak power is increased to about 0.5 GW. The relative bandwidth (FWHM) of the final output FEL spectrum is about $1.3 \times 10^{-4}$. Thus, the time-bandwidth product (TBP) describing the longitudinal coherence of the FEL radiation of the 30th harmonic can be calculated to be 0.649, which is only 1.47 times the Fourier-transform limit. Consequently, the resonance of the self-modulator tuned at the third harmonic is a favorable configuration for short-wavelength FEL at SXFEL-UF.
\begin{figure}[b]
	\includegraphics[width=0.4\textwidth]{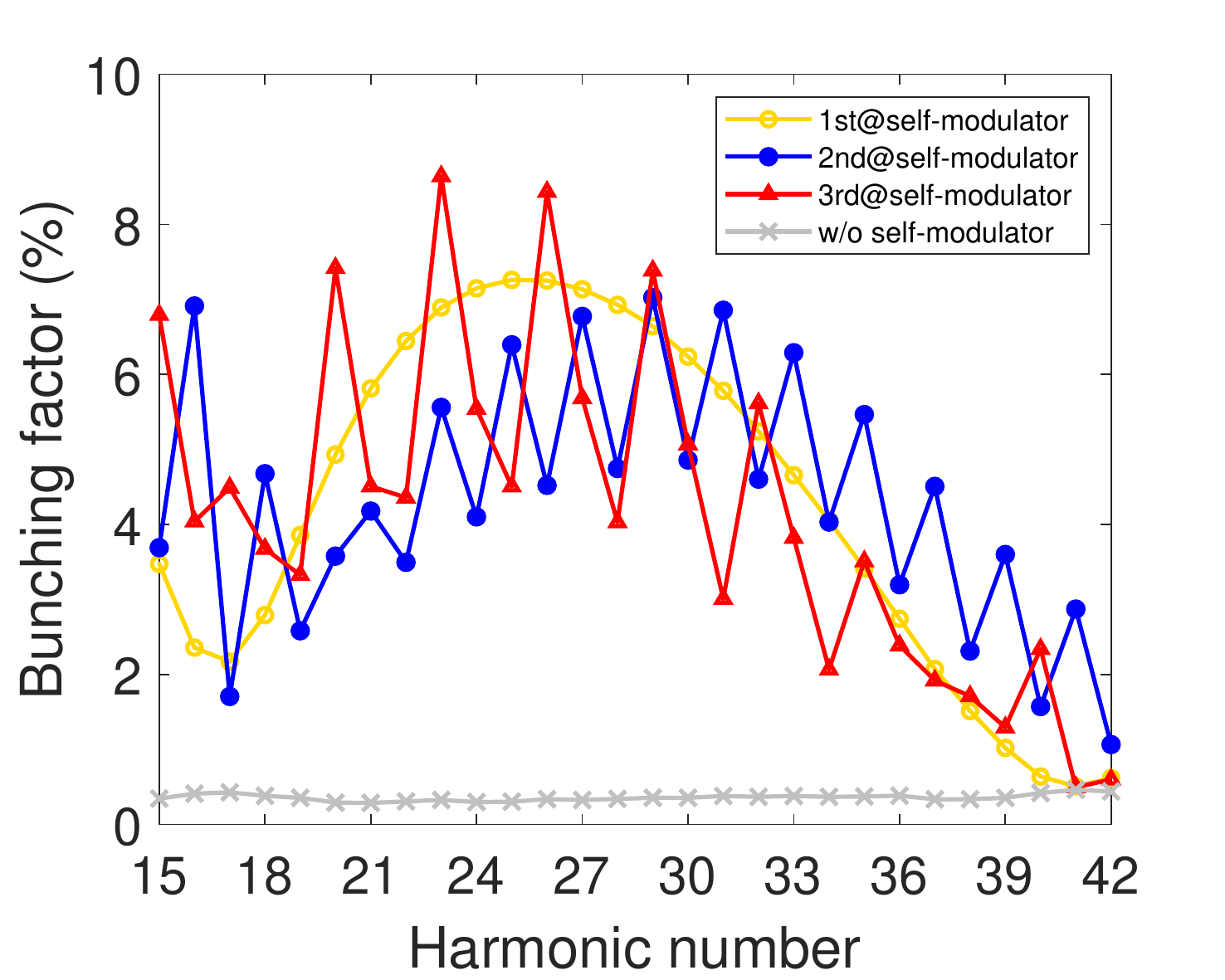}% Here is how to import EPS art
	\caption{\label{fig:10}Bunching factor after the second chicane as a function of the harmonic number in various cases, including without self-modulator and the resonance of the self-modulator tuned at the fundamental wavelength, second, and third harmonic of the seed laser, respectively.}
\end{figure}

\begin{figure*}[tb]
	\includegraphics[width=1\textwidth]{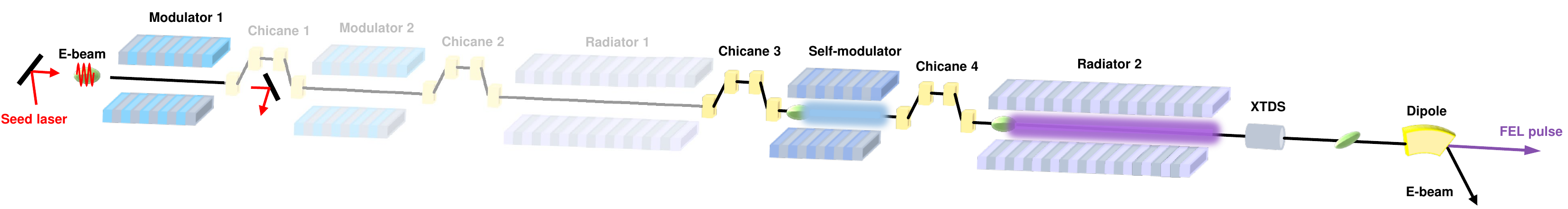}% Here is how to import EPS art
	\caption{\label{fig:11}The typical setup of the SXFEL-TF adopted a cascaded EEHG-HGHG scheme. In the self-modulation experiment, modulator 1, with a period of 80 mm in the first stage EEHG, was used as the first modulator. Chicane 3 is the fresh bunch chicane used as the first chicane. A modulator of the second stage HGHG with a period of 55 mm was the self-modulator. Chicane 4 was regarded as the second chicane.}
\end{figure*}

We further explore various harmonic self-modulation cases. The dispersion strength $R_{56}^2$ is optimized to obtain the maximum bunching factor at the 30th harmonic, while the bunching factor at various harmonic numbers are plotted in Fig.~\ref{fig:10}. In a standard HGHG, the bunching factor decreases exponentially as the harmonic number rises. The initial energy modulation of the first modulator is only 6 with tens of MW level external seed laser. If no self-modulator is introduced, the maximum bunching factor at the 30th harmonic is only $5 \times 10^{-5}$, just the shot noise level. Under the same initial energy modulation, the optimized maximum bunching factor at the 30th harmonic is 6.5\%, not the peak in the fundamental-wavelength case because of the additional energy spread. If the self-modulator resonates at the second harmonic, the enhanced coherent energy modulation can lead to bunching enhancement of the even harmonics. However, as shown in Fig.~\ref{fig:10}, the optimal bunching factor for the 30th harmonic is close to 5\% is not the peak. Instead, the 29th and 31st harmonics are enhanced, reaching almost 7\%. Similarly, in the third-harmonic self-modulation, the integer harmonics of 3 should be enhanced. The optimal optimized bunching factor at the 30th harmonic is close to 5\% but not the peak. The 29th and 31st harmonic bunching are enhanced to 7.5\% and 5.8\%, respectively. This evident frequency shift may cause by the same working points under different resonance conditions, such as almost identical dispersion strength $R_{56}$. Besides, the electron beam has been pre-modulated when it enters the self-modulator. The energy spread of the pre-bunched beam becomes larger while the energy distribution changes, and this process may also lead to a frequency shift. According to Fig.~\ref{fig:AvsBf}, if the initial energy modulation amplitude $A_1$ reaches 6, the pre-bunched beam entering the self-modulator has a more significant bunching factor at the second, third, or even higher harmonics. In brief, the self-modulation HGHG can generate an achievable bunching factor of nearly 5\% at the 30th harmonic reaching the soft x-ray wavelength range, where the standard HGHG is unattainable.

\section{\label{sec:sec5}Experiments}
The proof-of-principle experiments were conducted in SXFEL to demonstrate the self-modulation scheme. In the previous experiment \cite{Yan2021}, an electron beam with a laser-induced energy modulation as small as 1.8 times the slice energy spread was employed and the initial energy modulation was enhanced by three folds through the self-modulation method. Then, the electron beam was used for lasing at the 7th harmonic of a 266-nm seed laser in a single-stage HGHG setup and 30th harmonic of the seed laser in a two-stage HGHG setup.

Here, we further experimentally demonstrate the change of the resonance of the self-modulator can amplify the laser-induced energy modulation and generate a high harmonic bunching. The experiment was conducted at the SXFEL test facility (SXFEL-TF) \cite{Liu2021}, which employs cascaded EEHG-HGHG scheme with the “fresh bunch” technique. The schematic layout of the SXFEL-TF is displayed in Fig.~\ref{fig:11}. In the experiment, we used the modulator 1 and fresh bunch chicane of the first stage EEHG and the modulator, dispersive section, and radiator of the second stage HGHG to form the self-modulation HGHG. The parameters developed are electron beam energy of 780 MeV, bunch charge of 550 pC, peak current of 600 A, normalized emittance of 1.5 mm$\cdot$mrad, and envelope size of 300 $\mu$m. The pulse length and wavelength of the seed laser were 160 fs (FWHM) and 266 nm, respectively.

As shown in Fig.~\ref{fig:11}, the external seed laser was used to modulate the electron beam at the modulator 1. The dispersion strength of the first chicane was set to 0.24 mm. At first, the self-modulator was removed. The first undulator segment of the radiator 2, which can contain resonances from the 4th to more than the 11th harmonic of the seed laser, was used to produce coherent radiation and thus estimate the energy modulation level. The undulator period is 40 mm. For the standard HGHG case, a relative high power seed laser is used, where the coherent radiation at more than 6th harmonic can be detected. To ensure a weak seed laser power in the experiment, we continuously attenuated the seed laser intensity until the 4th harmonic signal could not be detected. In this case, the energy modulation amplitude was estimated as below four times the slice energy spread. 
\begin{figure}[bt]
	\includegraphics[width=0.4\textwidth]{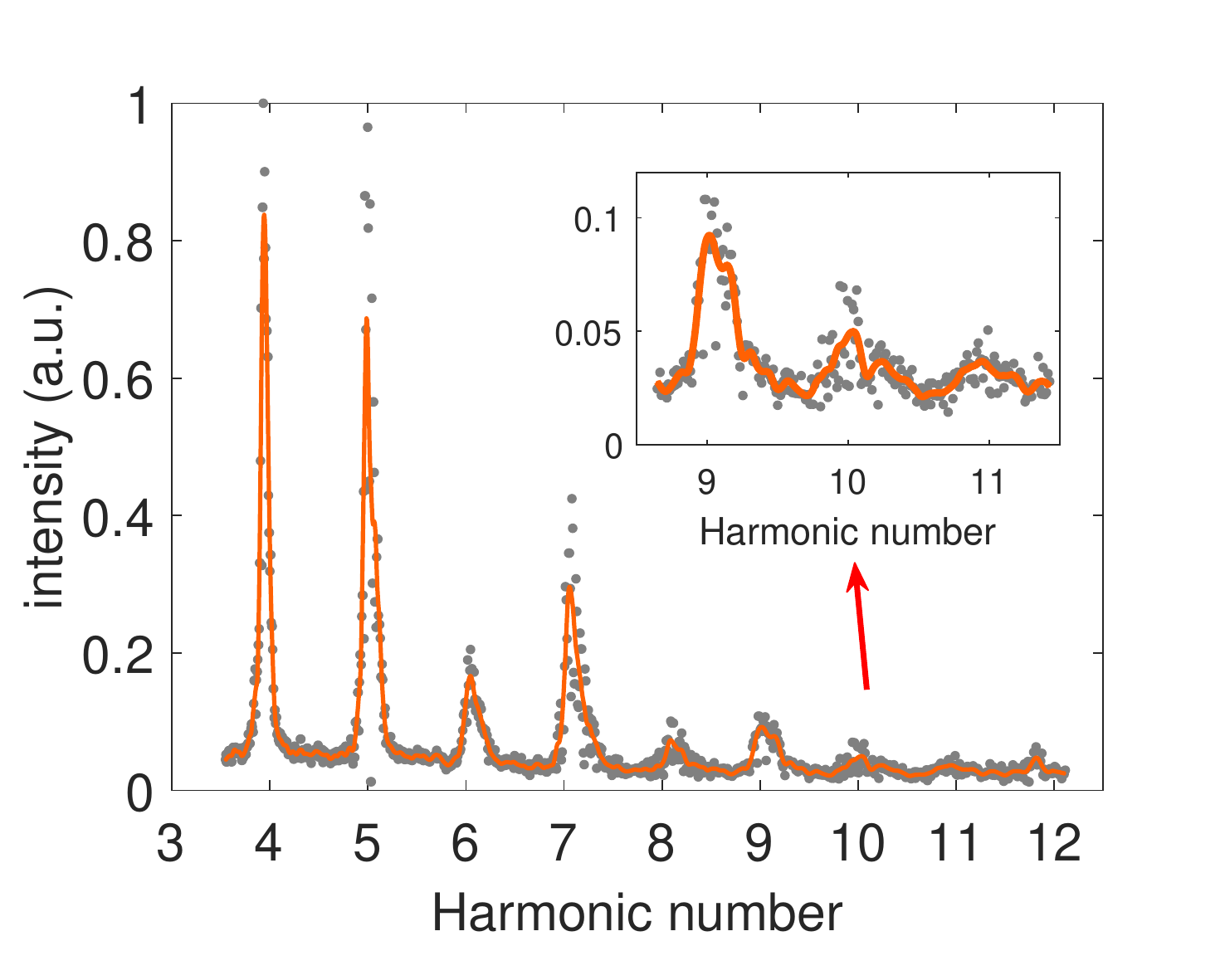}% Here is how to 
	\quad
	\includegraphics[width=0.4\textwidth]{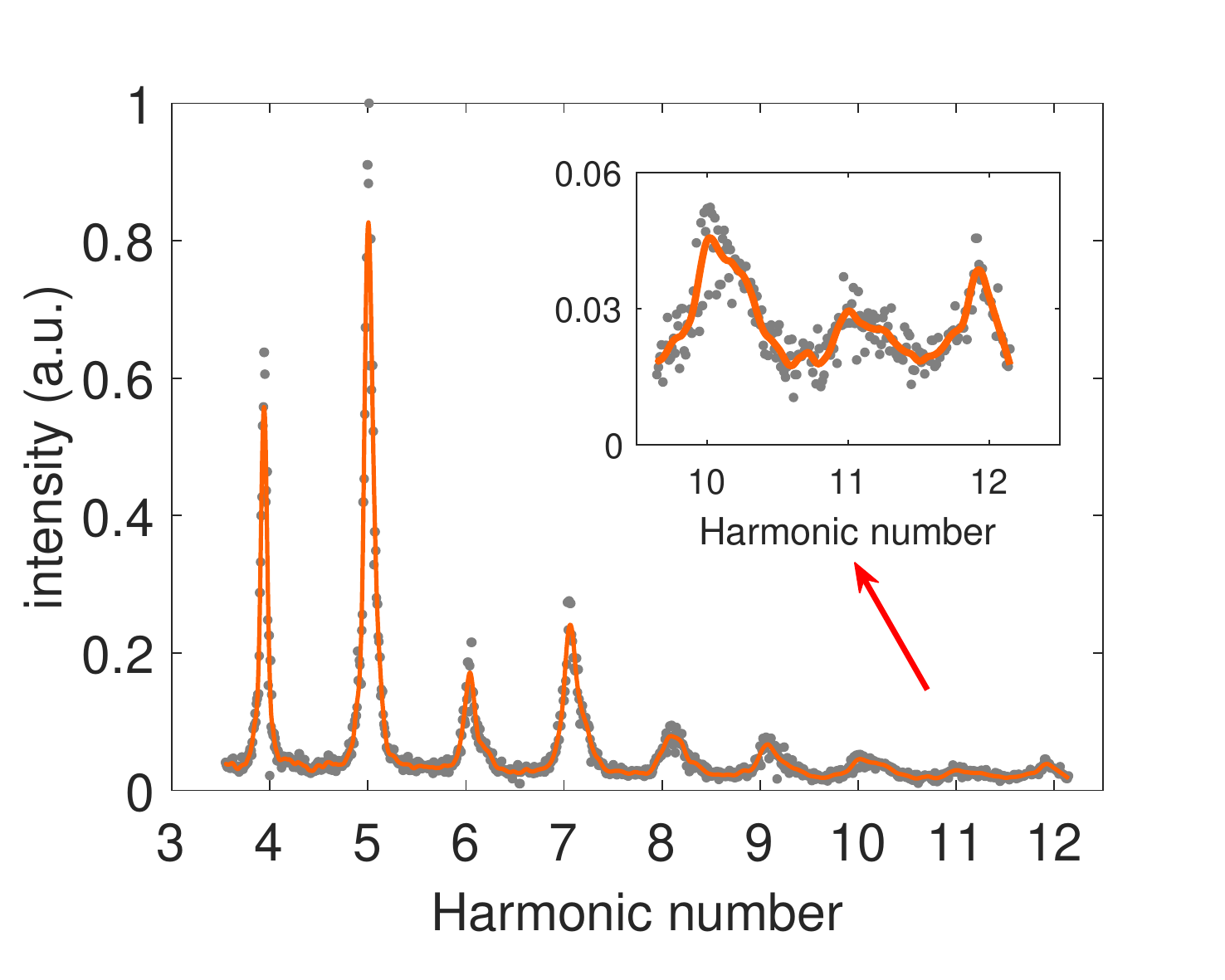}% Here is how to 
	\caption{\label{fig:12}The measured intensity of the coherent radiation at various harmonic numbers in the first undulator segment of the radiator, under different $R_{56}$ values of the second chicane of 0.038 mm (the top) and 0.048 mm (the bottom), respectively. The points represent the measurement results, and the curve represents the envelope obtained by smoothing the measurement data.}
\end{figure}

Subsequently, the resonance of the self-modulator was tuned at the second harmonic of the seed laser. The dispersion strength of the second chicane was set to a reasonable range to obtain a large harmonic bunching factor. Then, we continuously scanned the gap of the undulator segment. To enhance the radiation intensity at the target wavelength, we precisely optimized the dispersion strength of the second chicane to obtain the optimal $R_{56}$. Figure~\ref{fig:12} displays the measured coherent radiation intensity at various harmonic numbers, in which the dispersion strength of the second chicane was set to the optimal value of 0.038 mm at the 7th harmonic. The coherent radiation can be detected at up to the 11th harmonic, i.e., 24.2 nm. Furthermore, we slightly increased the dispersion strength of the first chicane to 0.28 mm. Then, the coherent radiation can be detected at even the 12th harmonic under the dispersion strength of 0.048 mm, as shown in Fig.~\ref{fig:12}. Since the external seed laser does not directly introduce the coherent energy modulation, and the self-modulator resonates at the second harmonic, the coherent radiation-based method is challenging to measure the energy modulation amplitude $A_2$ at the exit of the self-modulator through the coherent radiation-based method \cite{Feng2011}. Compared with the previous experiment (see Fig. 3 (a) in Ref. \cite{Yan2021}), the bunching performance is better than the case with an energy modulation amplitude of six times the slice energy spread enhanced by a 266-nm radiation.

In addition, we notice that the frequency mixing phenomenon experimentally, where the bunching factor at even harmonics is not significantly enhanced compared to the odd ones, as shown in Fig.~\ref{fig:12}. Because the initial energy modulation at the fundamental wavelength introduces a relatively large energy spread, the energy distribution of the electron beam is changed. After the harmonic self-modulation process, the energy spread increases, and the energy distribution of the electron beam is changed again while generating a high-frequency component that may lead to the frequency shift. Moreover, the dispersion strength of the first chicane may not be optimal, resulting in a weak harmonic energy modulation in the self-modulator. The bunching at the fundamental wavelength of the pre-bunched beam is maintained, which in combination with the energy modulation of the second harmonic leads to frequency mixing. In brief, the experiment results prove that the resonance of the self-modulator tuned at the second harmonic of the seed laser can still amplify the laser-induced energy modulation and generate a high harmonic bunching. The harmonic self-modulation scheme holds the promise to realize fully coherent soft x-ray FELs by reasonably optimizing the electron-beam orbit, the initial weak energy modulation, the resonance condition of the self-modulator, and the dispersion strength of chicanes.

\section{\label{sec:sec6}Summary and Prospects}
In summary, we systematically analyze and optimize the self-modulation scheme for further application in seeded FELs. The effects of critical parameters such as electron beam size, peak current, and relative length of the self-modulator are discussed in detail. The numerical simulations using the parameters of the SXFEL show that the self-modulation scheme can relax the requirement of the seed laser power by around three orders of magnitude. Moreover, the self-modulation HGHG is promising to lase at high harmonics of the seed laser through further optimizing the initial energy modulation and changing the resonance condition of the self-modulator. The numerical simulation has demonstrated that the self-modulation HGHG can lase at the 30th harmonic of the seed laser. In addition, we experimentally demonstrate that the harmonic self-modulation scheme can amplify the laser-induced energy modulation and generate a high harmonic bunching. Experimental results show that changing the resonance of the self-modulator at the second harmonic can still effectively amplify coherent energy modulation, in which case coherent radiation generated at up to the 12th harmonic can be observed.

The self-modulation scheme is promising to realize high-repetition-rate seeded FELs and those FEL schemes that require a higher power laser \cite{Allaria2007,Garcia2016,Zhou2017}. This scheme also promises to enable light sources with short wavelengths but low power, such as nonlinear harmonic generation \cite{Lembert2008,Labat2011,Ackermann2013}, to be used as seeds for the seeded FEL. Moreover, the harmonic self-modulation has the potential to produce MHz-repetition-rate and fully coherent radiation at a high harmonic of the external laser, significantly broadening the prospects of FEL applications.

\begin{acknowledgments}
The authors would like to thank N. S. Huang, Z. F. Gao, and W. J. Fan for their helpful discussions and comments. This work was supported by the National Key Research and Development Program of China (2018YFE0103100), the National Natural Science Foundation of China (12125508, 11935020), Program of Shanghai Academic/Technology Research Leader (21XD1404100), and Shanghai Pilot Program for Basic Research - Chinese Academy of Science, Shanghai Branch (JCYJ-SHFY-2021-010).
\end{acknowledgments}

\nocite{*}

\bibliography{manuscript}% Produces the bibliography via BibTeX.

\end{document}